\documentclass[letter]{aa}
\usepackage{graphicx}
\usepackage{natbib}
\usepackage{txfonts}
\usepackage{color}
\usepackage{hyperref} % I like to see references in a different colour
\hypersetup{
  colorlinks=true,
  pdfborder=2 2 0.,
  linkcolor=blue,
  anchorcolor=black,
  citecolor=blue,
  urlcolor=black,
}
\usepackage[normalem]{ulem}  % Used for strikeout text \sout, can be removed after
\usepackage{booktabs}
\bibpunct{(}{)}{;}{a}{}{,}

\usepackage{subfig}

\begin{document}

  \title{Gaia DR2 orbital properties for field stars with globular cluster-like CN band strengths}
  
  \author{A. Savino\inst{1} \and L. Posti\inst{1,2}}

\institute{Kapteyn Astronomical Institute, University of Groningen, Postbus 800, 9700 AV Groningen, The Netherlands.
            \email{A.savino@rug.nl}
           \and Observatoire Astronomique, Universit\'e de Strasbourg, CNRS, 11, rue de l`Universit\'e, F-67000 Strasbourg, France. \email{lorenzo.posti@astro.unistra.fr}
           }

 \abstract{Large spectroscopic surveys of the Milky Way have revealed that a small population of stars in the halo  have light element abundances comparable to those found in globular clusters. The favoured explanation for the peculiar abundances of these stars is that they originated inside a globular cluster and were subsequently lost.}
{Using orbit calculations we assess the likelihood that an existing sample of 57 field stars with globular cluster-like CN band strength originated in any of the currently known Milky Way globular clusters.}
{Using Sloan Digital Sky Survey and Gaia data, we determine orbits and integrals of motion of our sample of field stars, and use these values and metallicity to identify likely matches to globular clusters. The pivot hypothesis is that had these stars been stripped from such objects, they would have remained on very similar orbits.}
{We find that $\sim 70\%$ of the sample of field stars have orbital properties consistent with the halo of the Milky Way; however, only 20 stars have likely orbital associations with an existing globular cluster. The remaining $\sim 30\%$ of the sample have orbits that place them in the outer Galactic disc. No cluster of similar metallicity is known on analogous disc orbits.}
{The orbital properties of the halo stars seem to be compatible with the globular cluster escapee scenario. The stars in the outer disc are particularly surprising and deserve further investigation to establish their nature. }
\keywords{}
\authorrunning{A. Savino \& L. Posti}
\titlerunning{Orbital properties of CN-strong stars}
  \maketitle

%_____________________________________________________________________

\section{Introduction}
One of the major challenges in modern stellar population studies is to understand the origin of multiple stellar populations that are found in globular clusters (GCs). This phenomenon consists of specific elemental abundance variations within a given cluster that are not easily accountable for in the framework of stellar nucleosynthesis and chemical enrichment  \citep[e.g.][and references therein]{Bastian18}. The typical chemical pattern consists of enhancement in He, N, and Na (sometimes Mg and Si), and depletion in C and O (sometimes Al). This distinctive pattern is so ubiquitous in massive GCs, and exclusive to them, that it is now considered a signature feature to include in the very definition of these stellar systems.

However,  in the last decade evidence has accumulated that a fraction of stars in the Milky Way (MW) stellar field present very similar elemental abundances to those observed in GCs \citep[e.g.][]{Martell11, Fernandez-Trincado17,Schiavon17}. As these chemical anomalies have been detected using a variety of low- and high-resolution spectral indicators, in the rest of this manuscript we will generally refer to these stars as `enriched'. As GCs are expected to lose mass trough processes like evaporation and tidal stripping \citep[e.g.][]{Baumgardt03}, the favoured hypothesis for the origin of enriched stars is that they were once part of a GC. However, this association has not yet  been confirmed unambiguously.

Understanding the formation mechanism of these chemically peculiar stars in the MW has two potential rewards. It is currently accepted that the chemical anomalies of multiple stellar populations arise only in the specific environment of a proto-GC where extremely high densities can be reached. This specificity is a stringent requirement of any formation scenario \citep[e.g.][]{Renzini15}. Establishing whether enriched stars in the MW field formed in a GC could either support the specificity requirement or reveal that multiple stellar populations can also form, to some extent,  in lower density environments. On the other hand, identifying their birthplace can provide important details on the assembly history of the MW, in particular on the role of GC dissolution in that history.

To this end, the dynamical characterisation of known enriched stars can provide  insight into their association with the GC system. Even though the spatial association among the components of a dissolving stellar system is quickly lost, kinematic coherence is much more long-lived and orbital parameters are not expected to change significantly for several Gyr \citep[e.g. ][]{Helmi99}. The first dynamical characterisation for a sample of enriched stars was performed by \citet{Carollo13} and \citet{Fernandez-Trincado17}. However, these studies employed ground-based proper motion measurements, resulting in relatively large uncertainties in the dynamical modelling. Thanks to the unprecedented precision reached by Gaia Data Release 2 (DR2) astrometric measurements \citep{GaiaBrown18}, we now have the chance to reconstruct the motion of enriched stars in much greater detail. This approach was taken by \citet{Tang19} on a sample of CN-strong  stars in the LAMOST survey. CN-strong stars are enriched stars that are identified through their strong absorption in the CN molecular band, thus suggesting enhancement in nitrogen.

In this paper we focus more closely on the suggested GC origin for these chemically atypical stars. We reconstruct the orbital properties for a sample of candidate enriched halo stars and compare them with a nearly complete sample of Galactic GCs, investigating the possible association. In \S~\ref{ch:data} we discuss our samples, in \S~\ref{ch:modelling} we describe our dynamical modelling, in \S~\ref{ch:results} we present our results, and we discuss the implications in \S~\ref{ch:conclusions}.

\section{Data and sample selection}
\label{ch:data}
\subsection{Enriched stars}
For our analysis, we start from the sample of CN-strong stars from \citet{Martell10} and \citet[hereafter M11]{Martell11}. This is a sample of 65 CN-strong stars observed in the SEGUE and SEGUE-2 surveys \citep{Yanny09}, selected to be metal-poor red giants. For these stars we recover the position in the sky, radial velocity, and metallicity from the Sloan Digital Sky Survey database, and other spectral and photometric parameters (see section \ref{ch:dist})

We cross-correlate this sample with the Gaia DR2 database to obtain parallax and proper motion information. After the cross-correlation, and accounting for stars in common between the SEGUE and SEGUE-2 sample, we end up with 57 CN-strong stars for which we have six-dimensional phase-space information.

\subsubsection{Distance determination}
\label{ch:dist}

To perform a reliable dynamical modelling, accurate distances are needed for our CN-strong sample. Gaia parallax ($\varpi$) information can be used to get reliable distances; however, for this particular sample, simple estimators such as $1/\varpi$ or even more complex probabilistic approaches such as that of \citet{Bailer-Jones18}, which rely on disc-like priors for the MW stellar density distribution, are biased or simply not adequate.

\citet{Martell11} provides distances to these stars by matching the dereddened $(g-r)_0$ colour to theoretical models for red giant stars of the appropriate metallicity. However, theoretical predictions for the colour of these stars are affected by uncertainties in many of the stellar model ingredients. This, combined with the steepness of the red giant branch tracks in the Hertzsprung-Russell diagram, can lead to potentially high systematics in the distance determination.

We thus rederived the distances to the 57 CN-strong stars using a more sophisticated approach. We built upon the Bayesian framework laid out by \citet{Burnett10} and evaluated the probability distribution that each star has a  specific combination of distance, age, mass and metallicity based on the observed astrometric, spectroscopic and photometric properties.

As input observables, we use the dereddened $g_0$ magnitude and $(g-r)_0$ colour, and the spectroscopically derived $T_{\rm eff}$, $\log{(g),}$ and [Fe/H] values taken from the Sloan Digital Sky Survey database. We complement this information with the Gaia parallaxes. We use the alpha-enhanced BaSTI isochrones \citep{Pietrinferni06}, covering the metallicity range $\rm -3.62 \leq [Fe/H] \leq -0.29$, to estimate the multivariate Gaussian likelihood across our four-dimensional parameter space. The comparison between the expected and measured parallaxes takes into account the zero-point of $-0.029$ \citep{Lindegren18}.

We estimate the final posterior probability distribution by combining the likelihood with priors on the physical parameters of the star. We use the recalibrated Kroupa mass function from \citet{Aumer09} as mass prior, combined with the three-component prior on age, distance, and [Fe/H] from \citet{Burnett10}. The best-fit stellar parameters and the related uncertainties are calculated from the first- and second-order momenta of the probability distribution, respectively.

\begin{figure}
\centering
\includegraphics[scale=0.65]{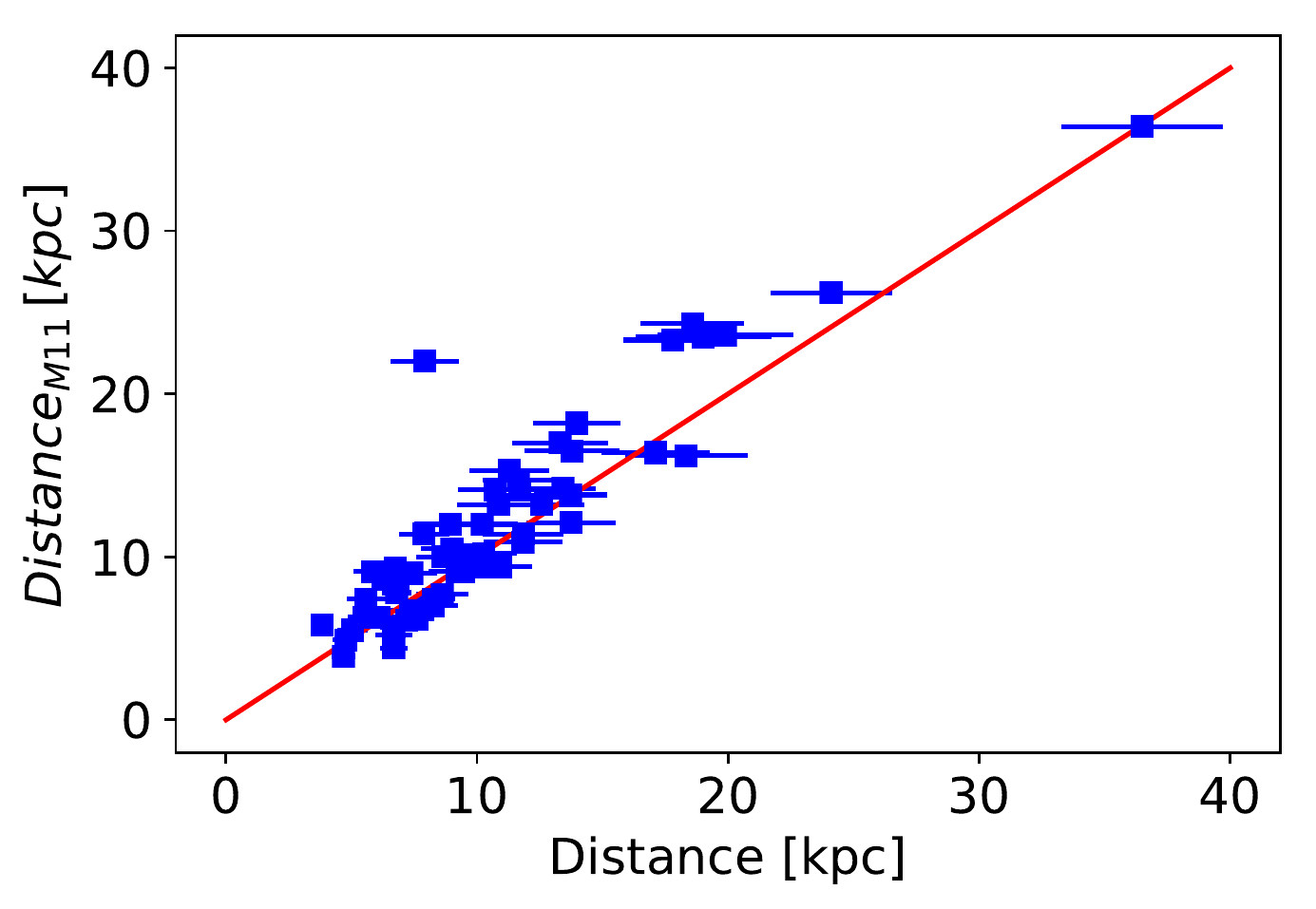}
\caption{ Comparison between our Bayesian distance inference and the distances from \citet{Martell11}. The solid line marks the 1:1 relation.}
\label{fig:D_comp}
\end{figure}

Figure~\ref{fig:D_comp} shows the comparison between our distance measurement and the M11 determination. Aside from one strong outlier (J224719.43+232144.9), the comparison is reasonable, with about 93\% of our sample having heliocentric distance differences below 4 kpc.

\subsection{Globular clusters}
For  the dynamical modelling of the GCs, we took the six-dimensional information from the catalogue of \citet{Vasiliev19}. This sample comprises 150 clusters; however,  we excluded Djorg 1, Terzan 10, and ESO456-78 because of their highly uncertain radial velocities. We assigned an uncertainty to cluster distances of 2.3\%, or 0.05 mag in distance modulus, as done in \citet{GaiaHelmi18}.

We adopted the cluster metallicities from \citet{Carretta09b}; for the 15 clusters not present in this study,  we took the metallicity from the  catalogue of \citet{Harris96} (2010 version). Two clusters are not in the \citet{Carretta09b} sample or the \citet{Harris96} catalogue:  Crater, for which we used the metallicity from \citet{Kirby15}, and 2MASS-GC03, for which we took metallicity and radial velocity from \citet{Carballo-Bello16}.

\begin{figure*}
\centering
\includegraphics[width=1.0\textwidth]{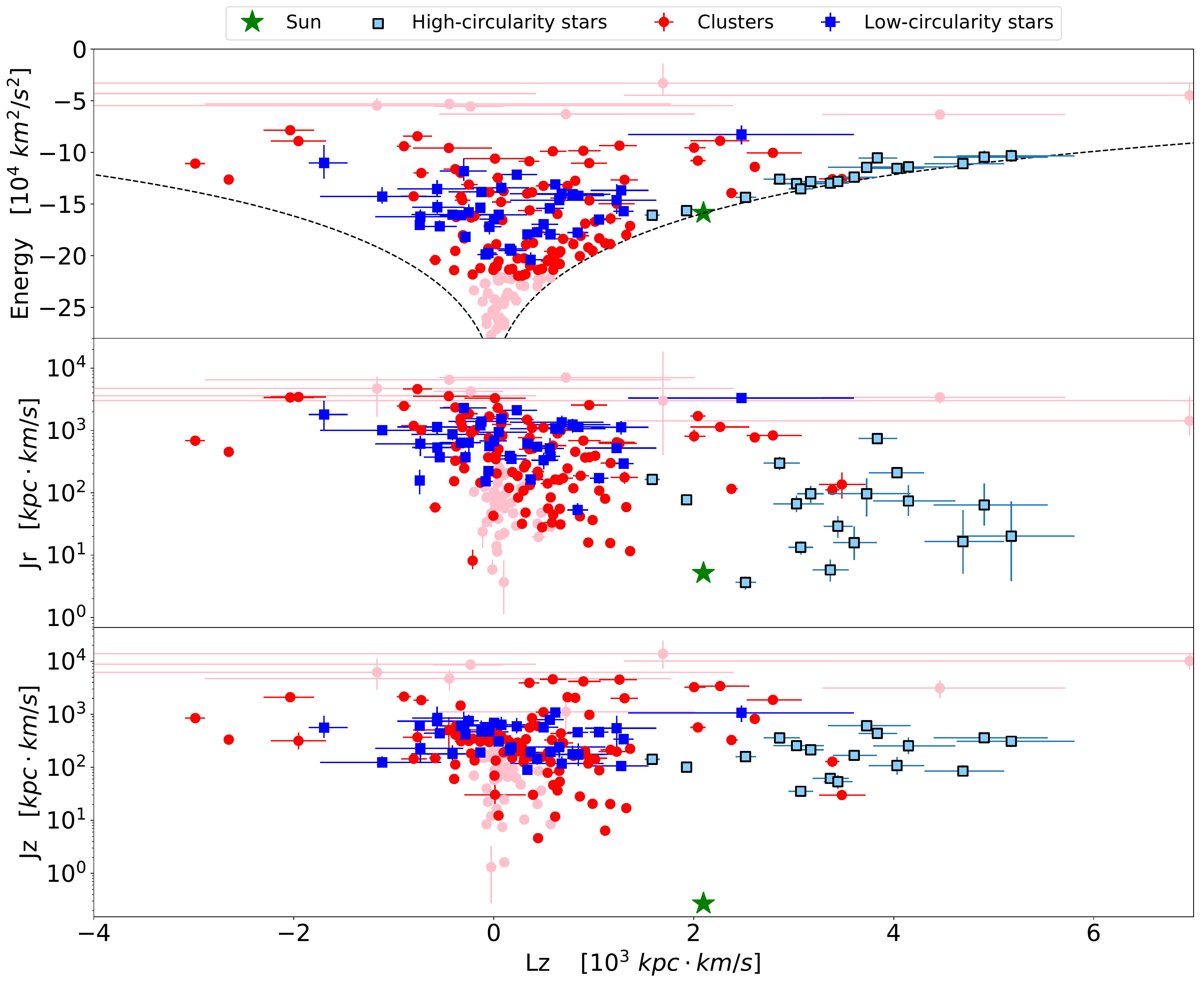}
\caption{Orbital energy (top), radial action (centre), and vertical action (bottom) as a function of the vertical angular momentum component for the globular clusters (red circles) and CN-strong stars (squares). Stars with circularity $>0.7$ are  shown in light blue and stars with circularity $\leq 0.7$ are  in dark blue. Clusters in pink are those with too high or too low orbital energy. The green star indicates the position of the Sun. The black dashed lines trace circular orbits. }
\label{fig:E_lz}
\end{figure*}

\section{Orbital modelling}
\label{ch:modelling}
We integrate orbits and compute integrals of motion for both the stars and the clusters using \texttt{AGAMA} \citep{Agama}. Some of the coordinate and velocity transformations make use of utilities from the \texttt{Galpy} \citep{Galpy} and \texttt{Astropy} \citep{Astropy12, Astropy18} packages. We adopt the MW potential from \citet{Piffl14}, a multi-component axisymmetric model fitted to the kinematics of stars in the solar vicinity, which is also compatible with the most recent determinations of the dark matter halo virial mass from Gaia DR2 \citep{Watkins18,Posti19}. The adopted galactocentric coordinates for the Sun are $X_{\odot} = 8.3$ kpc, $Y_{\odot} = Z_{\odot} = 0$ kpc. The circular velocity at the solar radius is $V_{\odot}^{\rm Circ} = 240.5$ km/s and for the peculiar velocity of the Sun we use ${\bf V}_{\odot}^{Pec} = [11.1,12.24,7.25]$ km/s \citep{Schonrich10}.

For each star and GC we calculate the orbital energy E, the vertical component of the angular momentum $L_z$, and the radial and vertical actions $J_r$ and $J_z$. The action integrals are computed via the `St{\"a}ckel Fudge' approximation \citep{Binney12}. We integrate the orbit backwards in time, with a time-step of 10 Myr, up to a look-back time of 5 Gyr.

Uncertainties on the orbital parameters are computed with a Monte Carlo approach. For each star and cluster, we extract 1000 realisations for the position and the velocity, randomly sampling the uncertainties in the distance, radial velocity, and proper motion\footnote{The uncertainties on the sky position are assumed to be negligible.}. The correlation between the two proper motions is taken into account, as is the systematic uncertainty of $0.035$ mas/yr \citep{GaiaHelmi18}. We thus construct the distributions of each quantity and associate its uncertainties by means of the 15.87 and 84.13 percentiles.

\section{Results}
\label{ch:results}

Figure~\ref{fig:E_lz} shows the integrals of motion for both the GC and CN-strong sample. The clusters span a much wider range of orbital energies than the CN-strong stars, thus we exclude  clusters with $E < -2.2\cdot 10^5 {\rm km^2/s^2}$ from the rest of our analysis. These clusters are confined in the inner regions of the Galaxy and are unlikely to be dynamically associated with our stellar sample. Similarly, also the distant clusters with $E > -7\cdot 10^4 {\rm km^2/s^2}$ can be excluded\footnote{Moreover, for these clusters the phase-space coordinates, and thus the integrals of motion, are very uncertain.}. The excluded clusters are shown as the pink  circles in Fig.~\ref{fig:E_lz}.

While GCs smoothly occupy the region of low angular momentum orbits for a broad range of energies, the CN-strong stars seem to belong to two distinct orbital families. While most of the sample has integrals of motion compatible with halo kinematics (low $|L_z|$ and high $E$, $J_r$, $J_z$), another group of stars sits at relatively high $L_z$ for their energy. Their orbits are in fact very close to circular (dashed black line in Fig.~\ref{fig:E_lz}; see also their trajectories in Fig.~\ref{fig:orbits}), with typical galactocentric radii between 10 and 20 kpc and typical vertical excursions of a few kpc. We thus calculate the orbital circularity of the stars as
\begin{equation} \label{eq:circ}
    c \equiv \frac{L_z}{L_z^{\rm Circ}(E)},
\end{equation}
where $L_z^{\rm Circ}(E)$ is the angular momentum of the circular orbit at the same energy $E$ of the star, and divide our sample into low-circularity stars ($c \leq 0.7$, dark blue squares in Fig.~\ref{fig:E_lz}) and high-circularity stars ($c > 0.7$, light blue squares with black edges).
The high-circularity stars have  vertical actions and energies that are similar to the low-circularity stars, while having markedly higher angular momenta and smaller radial action, hence smaller eccentricities.

\begin{figure}
\centering
\includegraphics[width=0.5\textwidth]{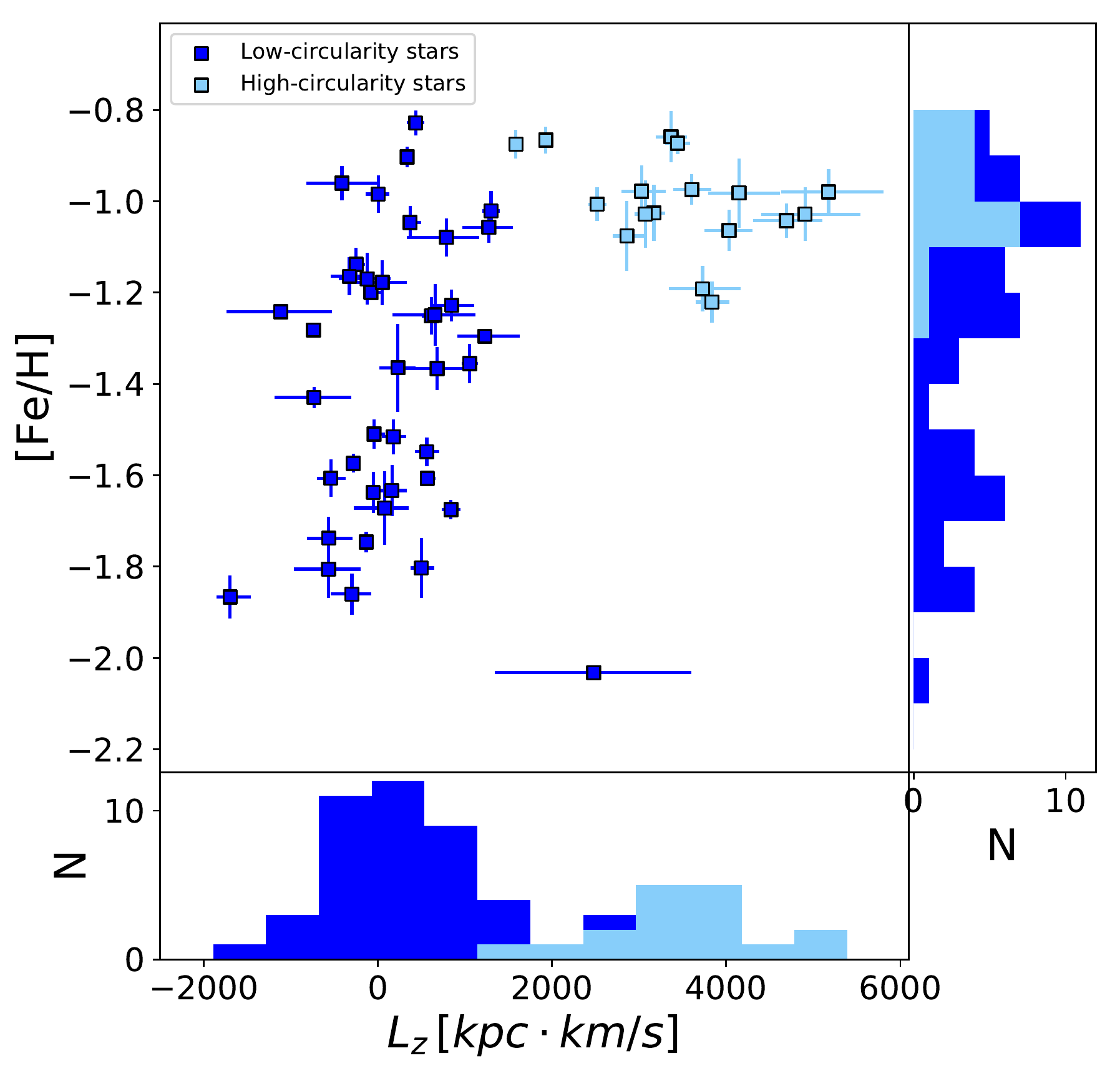}
\caption{Plot of  $L_z$ against metallicity for the CN-strong stars. Stars with circularity $> 0.7$ are shown in light blue, while stars with circularity $\leq 0.7$ are shown in dark blue. The side panels show the distribution on the respective axis of the whole sample and of the high-circularity stars.}
\label{fig:MDF}
\end{figure}

The 17 high-circularity stars, representing 29\% of our sample of CN-strong stars, also differ from the low-circularity stars in their chemical composition. Figure~\ref{fig:MDF} shows the $L_z$ and [Fe/H] distribution of our stellar sample. While the low-circularity stars are distributed roughly uniformly in metallicity, the high-circularity sample presents a very narrow metallicity distribution, peaking around -1.0. These stars in particular are found close to the upper limit of the metallicity selection adopted by M11, and thus we cannot exclude that we are seeing the low-metallicity tail of a larger, more metal-rich population. When comparing the kinematics of CN-strong stars with that of GCs, we find that the low-circularity stars lie in a region of the integrals of motion space that is compatible with some of the known MW GCs. Conversely, no cluster has orbital properties compatible with the high-circularity stars, with the exceptions of Pal1, E3, and ESO224-8. However, these clusters are significantly more metal rich ([Fe/H] = -0.51, -0.73, and 0.03, respectively) than our high-circularity sample, which means  we can exclude them as  the original birthplaces of these stars.

\begin{figure}
\centering
\includegraphics[width=0.5\textwidth]{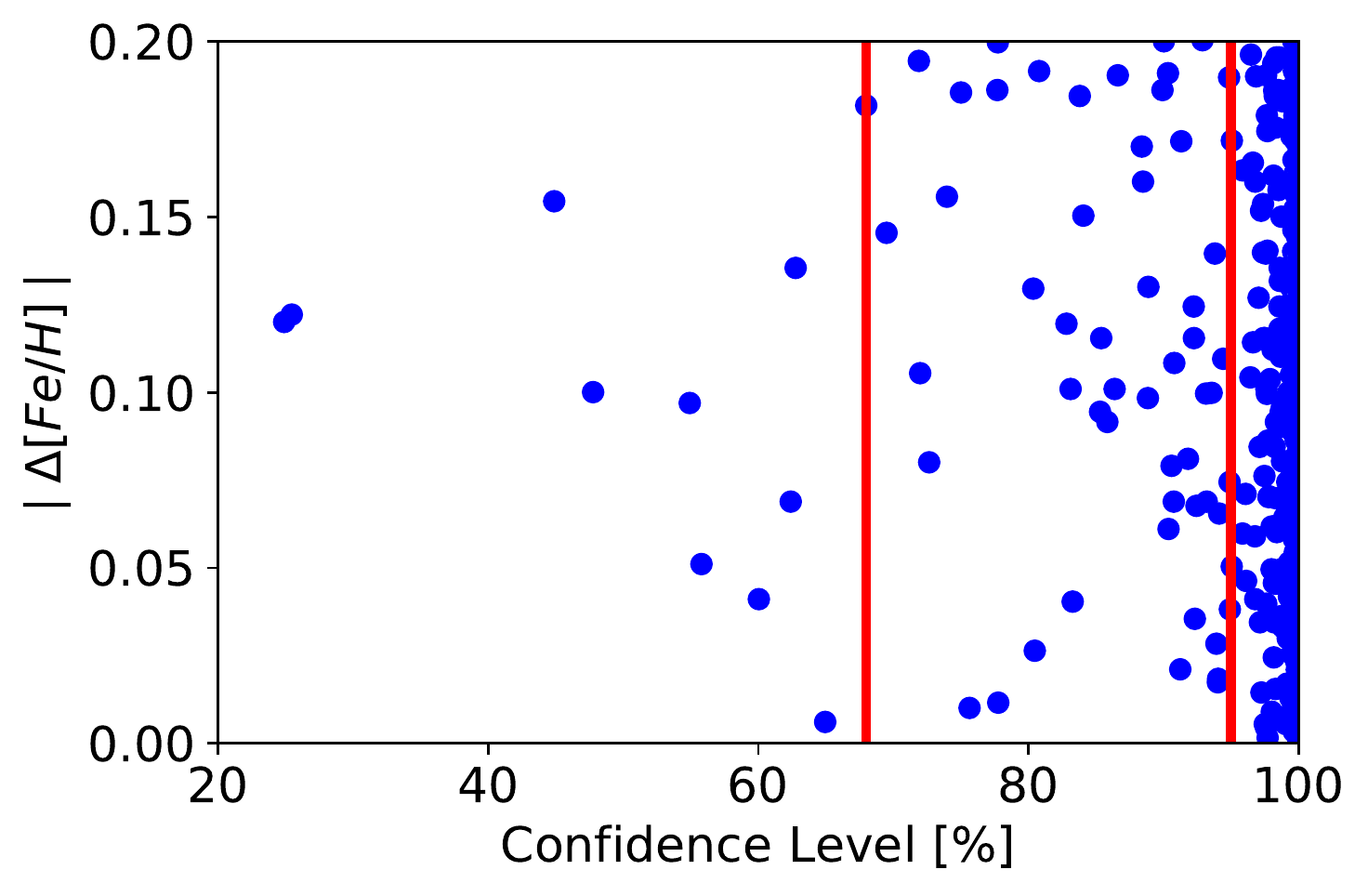}
\caption{For each cluster-star pair, confidence level for the pair to have different orbits against absolute difference in metallicity. Red lines indicate the 68\% and 95\% confidence level. Here the most likely star-cluster matches are those whose association can be rejected at a low confidence level.}
\label{fig:match}
\end{figure}

While Fig.~\ref{fig:E_lz} already shows that the low-circularity stars have orbits compatible with several GCs, the incredible precision with which the   DR2 data allow us to recover the integrals of motion provides the opportunity to perform a much more detailed and quantitative comparison. Since actions are adiabatic invariants, stars that have escaped a GC would still have  orbital actions that are very
similar to that of the cluster itself. Thus, we compared the values of $J_r$, $J_z$, and $L_z$ for every possible combination of CN-strong star and GC, totalling 5643 pairs. For each pair, we tested the null hypothesis that the cluster and the star have the same triplet of actions, taking into account the respective uncertainties. In Fig.~\ref{fig:match} we show, for each star-cluster pair, the degree of confidence at which the hypothesis of dynamical association can be rejected versus the absolute difference in [Fe/H]. In order to be escapees from a given cluster, a CN-strong star must have the same value of [Fe/H]. However, to accommodate observational uncertainties, as well as possible zero-point offsets, we consider every pair with an absolute difference in [Fe/H] less than 0.2 dex. The vast majority of orbital associations can be rejected at more than 95\% confidence level. Only 63 cluster-star associations with compatible metallicity cannot be excluded at this confidence level and, among them, only 11 are below the 68\% confidence level. These pairs are our best candidates for CN-strong star/GC association. We list the possible associations in table~\ref{tab:match}. We note that while for 9 stars we find a likely association with just one cluster, for 11 others the null hypothesis is not rejected for several clusters. In these cases we cannot unambiguously identify an association, but we can still conclude that GCs were likely the birthplaces for these stars.

We finally note that we  repeated our analysis using a different potential for the MW, as in \cite{McMillan17}. We find very similar results to those presented here, with 60 pairs identified at the 95\% confidence level.

\section{Discussion and conclusions}
\label{ch:conclusions}

In this letter we report on our  dynamical characterisation of a sample of CN-strong red giants; we also compared their properties with those of the Galactic GC system to test the likelihood that  these stars are cluster escapees.

We found that approximately one-third of our sample follows circular prograde orbits, with guiding radii of 10-20 kpc and vertical excursions of a few kpc. The orbital properties of the stars, and their relatively high metallicity, are much more similar to that of the Galactic thick disc, rather than to that of the stellar halo. The fact that some CN-strong stars from M11 are on disc-like orbits was already suggested by \citet{Carollo13}. However, the limited precision in their kinematic measurements prevented any stronger conclusion. The SEGUE footprint may play an important role in the selection of the outer thick disc envelope, and indeed Appendix~\ref{app:orb} shows that many high-circularity stars are observed at their largest height from the Galactic plane ($z_{\rm max}$). It is curious to note that these stars are mostly located towards the Galactic anticentre, which is where the Galactic disc appears to be highly vertically perturbed \citep[][]{Gomez16, Bergemann18, Laporte18}.

The kinematics of the high-circularity stars is extremely different from that of any known GC (with the exceptions discussed in \S~\ref{ch:results}). Given the available data, we cannot exclude that these stars have genuine GC-like abundances and trace a population of enriched stars unrelated to the current MW GC system. However, it is probable that these stars are contaminants in the selection of M11. At fixed metallicity and absolute magnitude (which are also known to affect the CN molecular feature used in M11), we note that the high-circularity stars are found to have the highest CN band strengths in our sample. This likely  indicates that they  have unusual compositions, and not that they are  interlopers from the chemically normal field population.

For the origin of the high CN absorption, mass transfer from an asymptotic giant branch companion is a commonly invoked mechanism \citep[e.g. ][]{Schiavon17}. Accretion from a massive ($3-8 \, M_{\odot}$) asymptotic giant branch star can enhance the atmospheric nitrogen abundance and mimic the light element pattern seen in GCs, but the fraction of field stars that undergo this enrichment process is expected to be low. Roughly two times the number of stars experience a similar process, with donors of $1.5-3 \, M_{\odot}$, and become CH-stars. These stars, characterised by high carbon abundance, also have strong CN absorption. As the selection from M11 is insensitive to carbon abundance at high metallicities, these stars are a likely source of contamination. Regardless of the donor mass, if binary mass transfer is the source of the chemical peculiarity in these stars, this could be confirmed by radial velocity monitoring. Unfortunately, the orbital periods of such binaries can be as long as many years \citep{Lucatello05,Starkenburg14}, requiring long observational campaigns. Alternatively, high-resolution spectroscopy can provide the detailed abundance patterns of both light and s-process elements and validate the asymptotic giant branch pollution scenario.

In this letter we discuss how we tested, for each CN-strong star, the hypothesis that it is dynamically associated with any of the GCs. We did this by comparing the three orbital actions and the metallicity. We excluded the vast majority of cluster-star associations with a confidence level above $95\%$. However, 63 associations could not be excluded with such confidence. In particular, 7 stars had  orbital and chemical properties that were very similar to those of at least one cluster, resulting in 11 promising star-cluster associations (see Table~\ref{tab:match}, in bold). These stars may be among the first GC escapees robustly identified, not considering tidal structures observed around a few GCs. For this reason, we encourage high-resolution follow-up observations to confirm the chemical abundances.

The fact that 50\% of our low-circularity CN-strong stars do not seem to be associated with any cluster in our sample does not necessarily refute the GC origin. There are a number of possible scenarios that can accommodate our findings. It is possible that some of our stars come from GCs that are not present in our analysis. While our GCs sample is nearly complete,  it is certainly possible that we have not discovered some of the massive MW GCs, for example those lying beyond the highly extinct regions of the Galaxy. However, the number of still undiscovered clusters is expected to be very small \citep{Harris13}. 

A second scenario is that these stars come from long dissolved GCs. The idea of a population of GCs that have been completely dissolved by the MW tidal field gained support from the observation of strong mass loss in a few existing GCs \citep{Rockosi02, Odenkirchen03, Belokurov06} and it has found confirmation with the discovery of prominent, dynamically cold stellar streams in the halo of our galaxy \citep{Grillmair06, Ibata18, Shipp18}. Thanks to the wealth of information provided by Gaia DR2, several of these features have been detected \citep{Ibata19}, suggesting that GC dissolution may be a rather common phenomenon. Furthermore, additional destruction channels have been proposed, such as dissolution in the turbulent discs  of small galaxies, later accreted by the MW \citep{Kruijssen15}. Given the above arguments, GC destruction may play an important contribution to the population of CN-strong stars observed in the Galaxy. However, multiple populations arise only in GCs above a mass threshold of around $10^4 M_{\odot}$ \citep{Carretta10, Dalessandro14, Bragaglia17, Simpson17}. Even above this threshold, the extent of light element variation and the fraction of enriched stars become larger with increasing mass \citep{Milone17}. Hence, it is not clear whether cluster dissolution can be an efficient source of CN-strong stars in the MW field, given also that stars in more massive clusters are less easily stripped.

We also note an important caveat. In the case of the low-circularity stars, even if our dynamical analysis did not yield a conclusive association with any of the GCs, it is still possible that these stars were born in one of the GCs in our sample. This could be the case given the limitations of our dynamical approach, which assumes that there is  equilibrium and  that escapees retain orbital memory. The former might  not be applicable, for instance in the case of a dynamically important accretion event, which can change the integrals of motions significantly; the latter  does not happen if a star was expelled from the cluster at high velocity, for example after a three-body interaction.

% A third hypothesis is that some of the CN-strong stars were lost very early in the cluster life. Any subsequent event that changed the MW potential rapidly and substantially, like a major accretion, would have changed the integrals of motion for the star and the cluster in a different way, resulting in missed association with our method. We know that the MW is currently accreting the Sagittarius dwarf galaxy \citep{Ibata94}. However, the mass is this galaxy is likely too small to provide a significant potential perturbation (ref). On the other hand, we know that our galaxy experienced at least one major accretion event between redshift 1 and 3 \citep{Helmi18,Belokurov18}, with a second similar event suggested \citep{Kruijssen18}. Any CN-strong star lost before or shortly after such tumultuous phase must have experienced significant variation in its orbital properties and thus it is unlikely to be associated with its progenitor cluster. \textbf{Should we mention the potential perturbation from the LMC?}

In conclusion, even though we were able to  assign one or more candidate parent clusters only to half of the low-circularity stars, the orbital properties of this subsample are broadly consistent with those of the halo GC population, supporting escape from GCs as a major formation channel for these stars. However, the inconsistency of such a scenario with the properties of high-circularity stars poses interesting questions. We strongly encourage follow-up analysis on these stars to unambiguously determine whether they are interlopers from a population of disc binary stars or if they have genuine GC-like abundances. In either  case, the answer to this question will add precious information to the specificity hypothesis for GC multiple stellar population formation scenarios.

\begin{acknowledgements}
We wish to thank E. Balbinot, A. Helmi, D. Massari, and E. Tolstoy for the insightful discussions during the preparation of this manuscript. LP acknowledges financial support from a VICI grant from the Netherlands Organisation for Scientific Research (NWO) and from the {\it Centre National d'Etudes Spatiales} (CNES).

\end{acknowledgements}

\bibliographystyle{aa}
\bibliography{Bibliography}

\clearpage

\appendix
\onecolumn
\centering
\section{Candidate star-cluster associations}

\begin{table*}[h!]\scriptsize
\renewcommand{\arraystretch}{01}
\centering
\caption{Possible star-cluster associations based on the orbital actions and metallicity. The columns list the star and cluster IDs, the degree of confidence for the orbital association rejection, and the absolute difference in metallicity.  Entries in bold are the associations that can be excluded at less than 68\% confidence level.}
\begin{tabular} {llcc}
\toprule
Star ID & Cluster & Confidence level & ${\mid\Delta[Fe/H]\mid}$\\
\midrule
J115957.53+044724.6  & NGC1261 & 85.3\% & 0.095\\
J115957.53+044724.6  & NGC1851 & 83.8\% & 0.185\\
J115957.53+044724.6  & NGC6229 & 94.1\% & 0.066\\
J115957.53+044724.6  & NGC6864 & 94.9\% & 0.075\\
J115957.53+044724.6  & NGC6981 & 92.3\% & 0.116\\
 \hline
J224241.48+131459.2  & NGC362  & 88.9\% & 0.130 \\
\textbf{J224241.48+131459.2}  & \textbf{NGC1261} & \textbf{47.8\%} & \textbf{0.100}  \\
J224241.48+131459.2  & NGC1851 & 75.6\% & 0.010\\
J224241.48+131459.2  & NGC5904 & 88.5\% & 0.160\\
\textbf{J224241.48+131459.2}  & \textbf{NGC6864} & \textbf{24.9\%} & \textbf{0.120}\\
 \hline
\textbf{J160327.56+173348.5}  & \textbf{NGC288} & \textbf{67.9\%} & \textbf{0.182}\\
J160327.56+173348.5  & NGC6723 & 94.9\% & 0.038\\
 \hline
J083423.50+124324.4  & NGC1851 & 77.7\% & 0.186\\
J083423.50+124324.4  & NGC2808 & 89.9\% & 0.186\\
\hline
J115826.53+043047.8  & NGC1904 & 85.8\% & 0.092\\
J115826.53+043047.8  & NGC4147 & 90.1\% & 0.108\\
J115826.53+043047.8  & NGC5286 & 93.9\% & 0.028\\
J115826.53+043047.8  & ESO280-6 & 88.8\% & 0.098\\
J115826.53+043047.8  & NGC6584 & 91.3\% & 0.172\\
J115826.53+043047.8  & NGC6981 & 80.8\% & 0.192\\
J115826.53+043047.8  & NGC7089 & 77.8\% & 0.012\\
J115826.53+043047.8  & NGC7492 & 94.0\% & 0.018\\
\hline
J162242.14+315035.7  & NGC6205 & 80.5\% & 0.026\\
\hline
J085455.54+372927.9  & NGC1261 & 90.3\% & 0.191\\
J085455.54+372927.9  & NGC1851 & 83.1\% & 0.101\\
J085455.54+372927.9  & NGC2808 & 86.4\% & 0.101\\
\hline
J161013.43+052403.6  & NGC6723 & 93.2\% & 0.100\\
\hline
\textbf{J102826.25+175448.9}  & \textbf{NGC362} & \textbf{25.4\%} & \textbf{0.122}\\
\hline
J161943.91+163110.9  & NGC288 & 94.9\% & 0.190\\
J161943.91+163110.9  & NGC6284 & 77.7\% & 0.199\\
\hline
J105019.09+005105.6  & NGC5904 & 90.6\% & 0.079\\
\hline
J152635.55+510607.6  & NGC288 & 74.0\% & 0.156\\
\hline
\textbf{J085342.07+364723.9}  & \textbf{NGC362} & \textbf{55.8\%} & \textbf{0.051}\\
J085342.07+364723.9  & NGC1261 & 91.2\% & 0.021\\
J085342.07+364723.9  & NGC1851 & 93.2\% & 0.069\\
\textbf{J085342.07+364723.9}  & \textbf{NGC2808} & \textbf{62.4\%} & \textbf{0.069}\\
J085342.07+364723.9  & NGC5904 & 91.8\% & 0.081\\
J085342.07+364723.9  & NGC6121 & 90.8\% & 0.069\\
J085342.07+364723.9  & NGC6284 & 90.4\% & 0.061\\
\textbf{J085342.07+364723.9}  & \textbf{NGC6864} & \textbf{60.0\%} & \textbf{0.041}\\
\hline
J014924.75+150045.4  & NGC288 & 94.4\% & 0.110\\
J014924.75+150045.4  & NGC362 & 80.4\% & 0.123\\
J014924.75+150045.4  & NGC5946 & 93.8\% & 0.140\\
J014924.75+150045.4  & NGC6205 & 84.1\% & 0.150\\
J014924.75+150045.4  & NGC6284 & 82.8\% & 0.120\\
J014924.75+150045.4  & NGC6544 & 83.3\% & 0.040\\
J014924.75+150045.4  & NGC6681 & 86.6\% & 0.190\\
\hline
J160709.22+044712.6 & NGC6681 & 94.0\% & 0.017\\
\hline
\textbf{J173052.92+333303.1}  & \textbf{NGC6205} & \textbf{65.0\%} & \textbf{0.006}\\
\hline
J172931.11+264940.2 & Pal15 & 92.4\% & 0.068\\
\hline
\textbf{J234549.70+005055.9}  & \textbf{NGC2298} & \textbf{44.9\%} & \textbf{0.155}\\
J234549.70+005055.9  & NGC5286 & 72.0\% & 0.106\\
J234549.70+005055.9  & NGC5634 & 92.2\% & 0.125\\
J234549.70+005055.9  & IC4499 & 75.0\% & 0.186\\
\textbf{J234549.70+005055.9}  & \textbf{IC1257} & \textbf{62.8\%} & \textbf{0.136}\\
J234549.70+005055.9  & ESO280-06 & 92.3\% & 0.036\\
J234549.70+005055.9  & NGC6779 & 71.9\% & 0.195\\
J234549.70+005055.9  & NGC7089 & 69.5\% & 0.146\\
J234549.70+005055.9  & NGC7492 & 85.4\% & 0.116\\
\hline
J125224.31+193358.3  & NGC2298 & 93.6\% & 0.100\\
J125224.31+193358.3  & NGC4147 & 72.7\% & 0.080\\
J125224.31+193358.3  & NGC7492 & 88.4\% & 0.170\\
\hline
\textbf{J131447.33+010356.4}  & \textbf{NGC5897} & \textbf{54.9\%} & \textbf{0.097}\\

\bottomrule
\end{tabular}
\label{tab:match}
\end{table*}
\clearpage

\section{Orbits of the CN-strong stars}
\label{app:orb}

\begin{figure*}[b!]
\captionsetup[subfigure]{labelformat=empty}
        \subfloat[][]
        {\includegraphics[width=\textwidth]{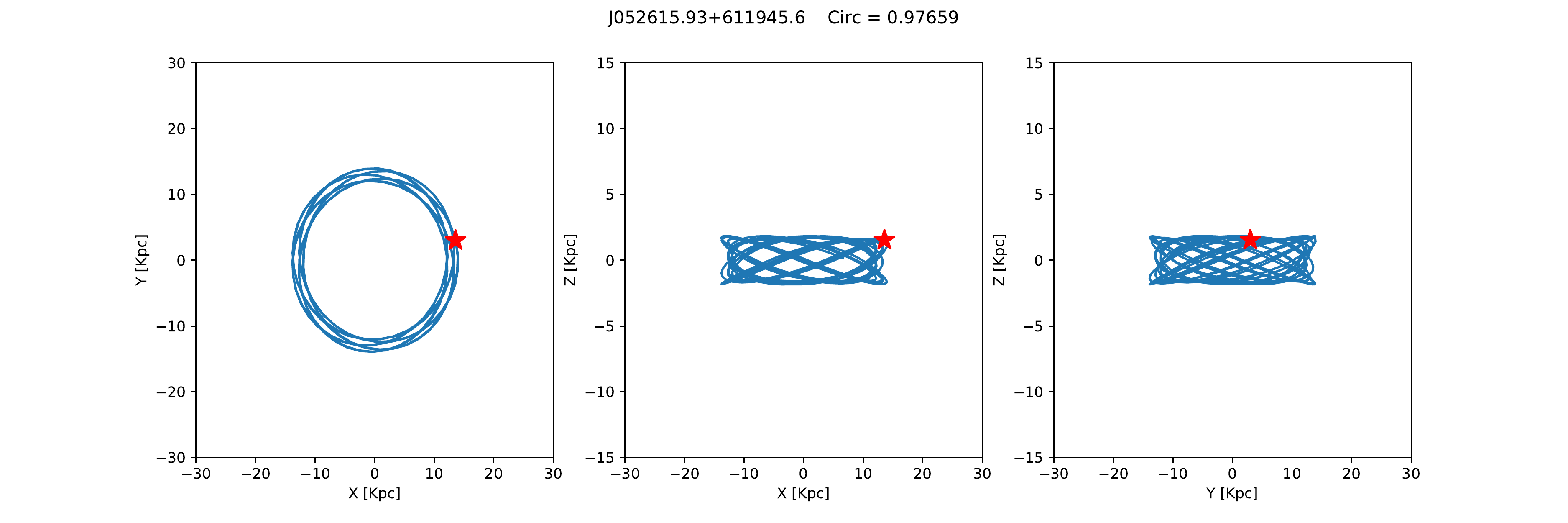}} \quad
            \subfloat[][]
        {\includegraphics[width=\textwidth]{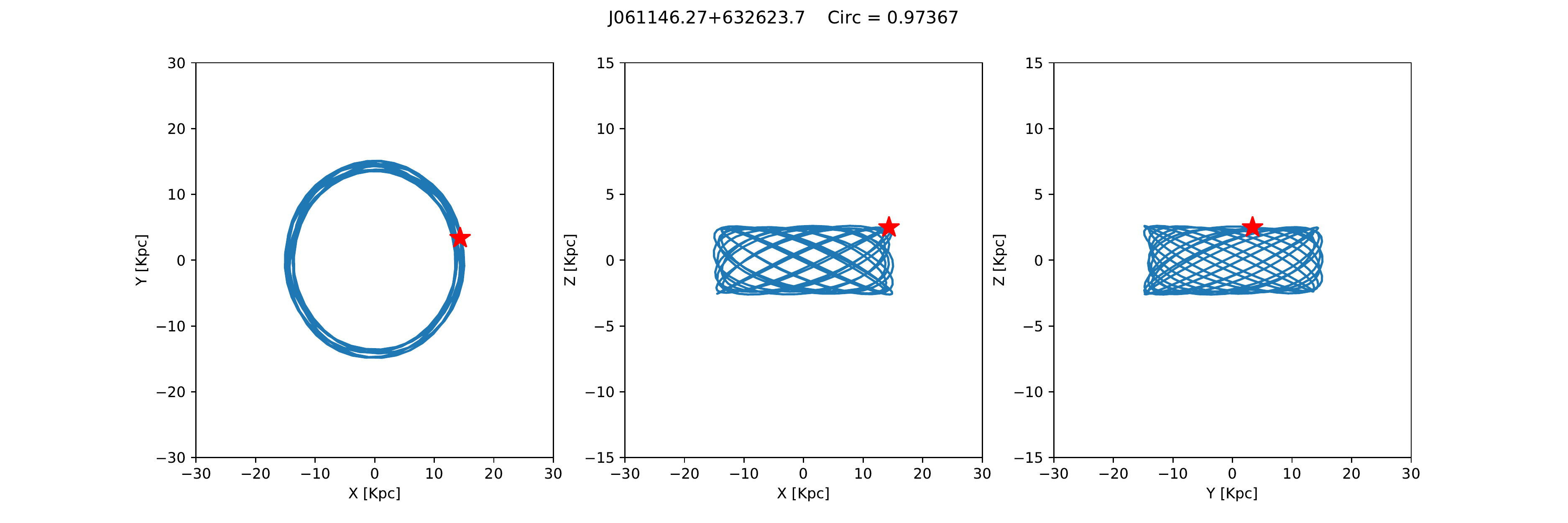}} \quad
            \subfloat[][]
        {\includegraphics[width=\textwidth]{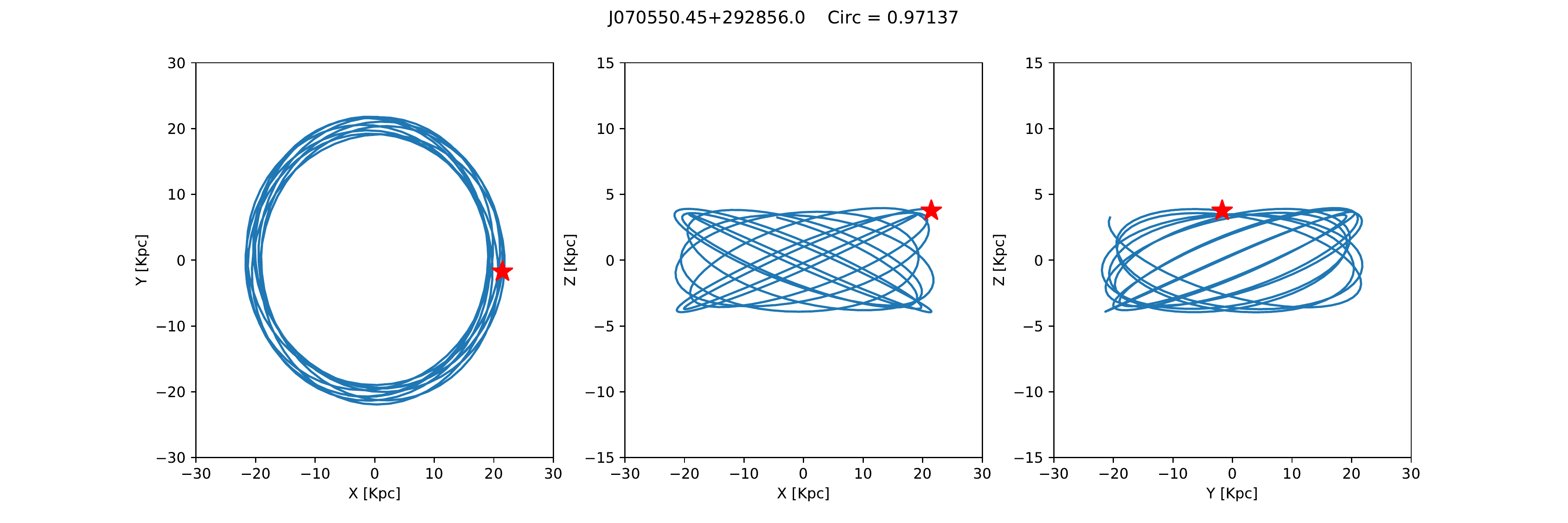}} \quad
        
    \caption{Projection, in the three galactocentric cartesian planes of the reconstructed orbit for the CN-strong stars used in this study, ordered by decreasing absolute value of circularity. The red star indicates the current position of the star in the Galaxy. The star ID and the circularity are given above each set of panels.}
    \label{fig:orbits}
\end{figure*}

\begin{figure*}\ContinuedFloat
\captionsetup[subfigure]{labelformat=empty}
        \subfloat[][]
        {\includegraphics[width=\textwidth]{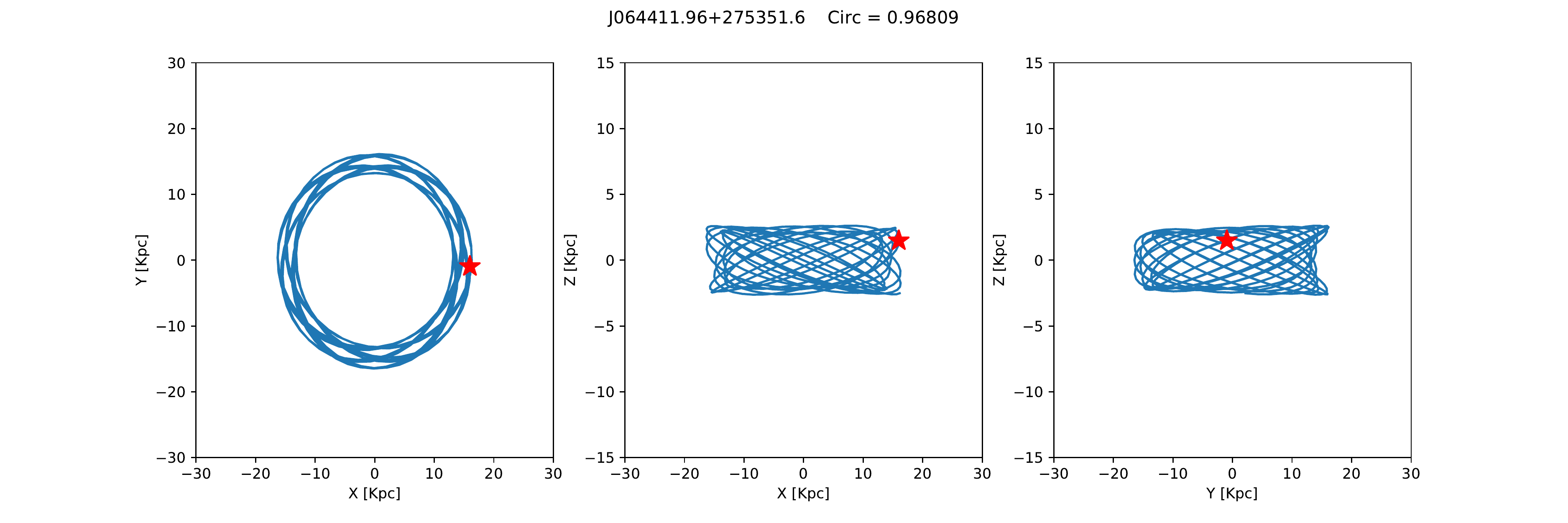}} \quad
            \subfloat[][]
        {\includegraphics[width=\textwidth]{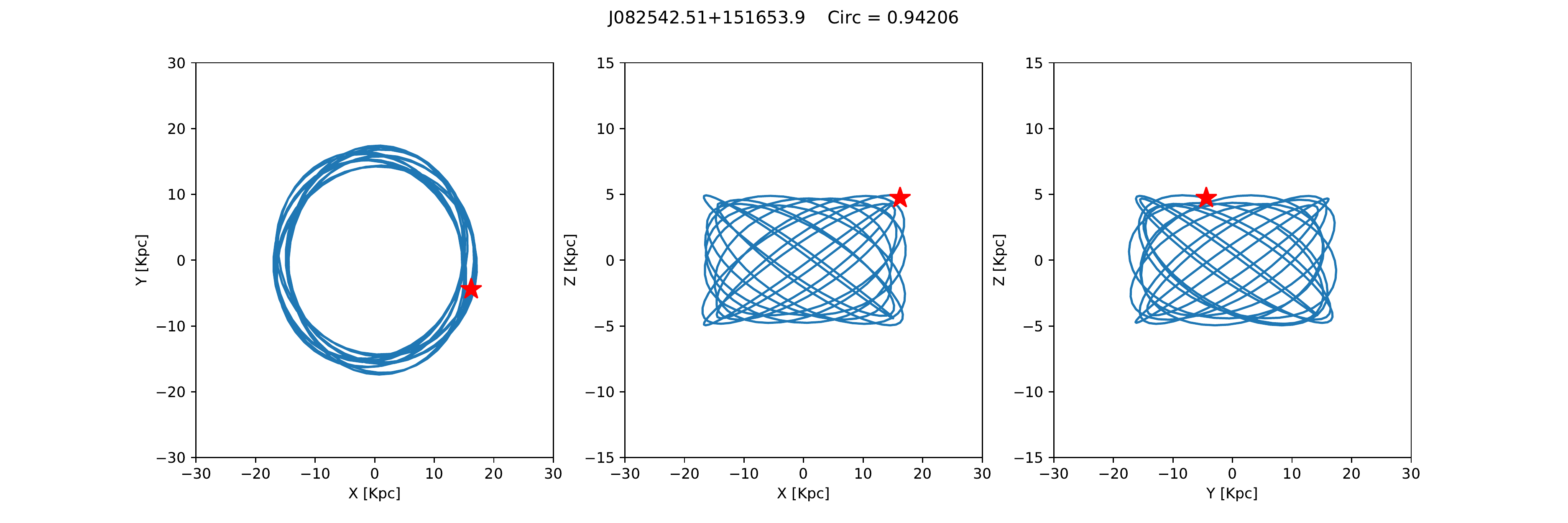}} \quad
            \subfloat[][]
        {\includegraphics[width=\textwidth]{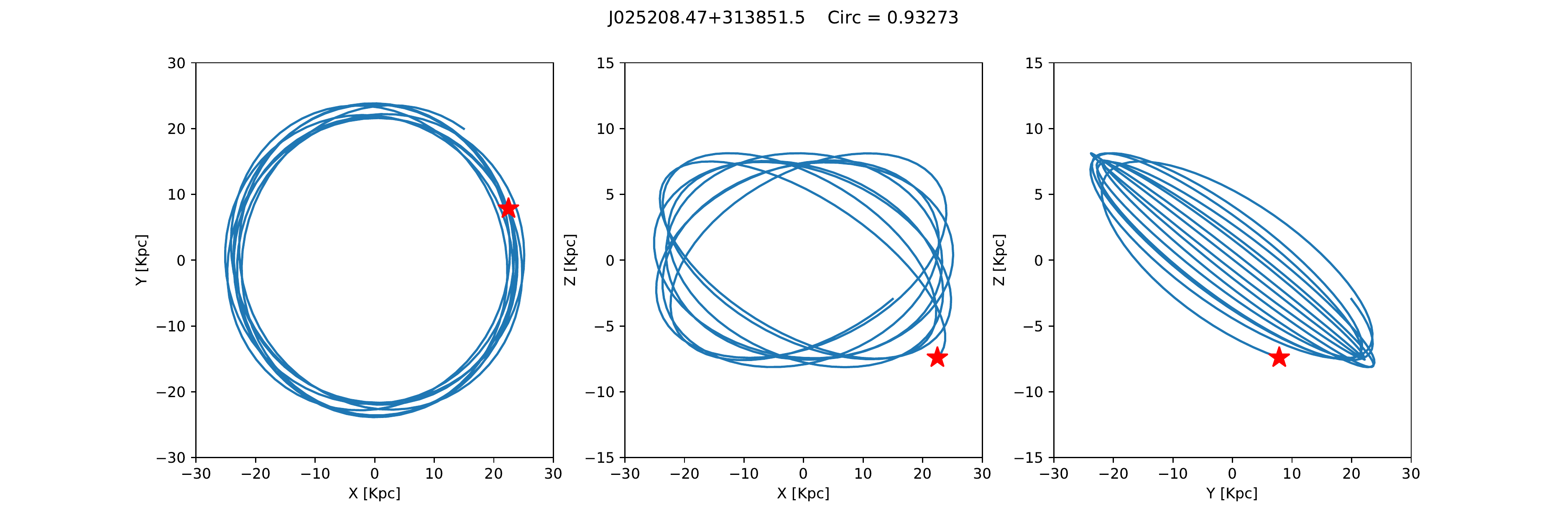}} \quad
        
    \caption{Continued.}
\end{figure*}

\begin{figure*}\ContinuedFloat
\captionsetup[subfigure]{labelformat=empty}
        \subfloat[][]
        {\includegraphics[width=\textwidth]{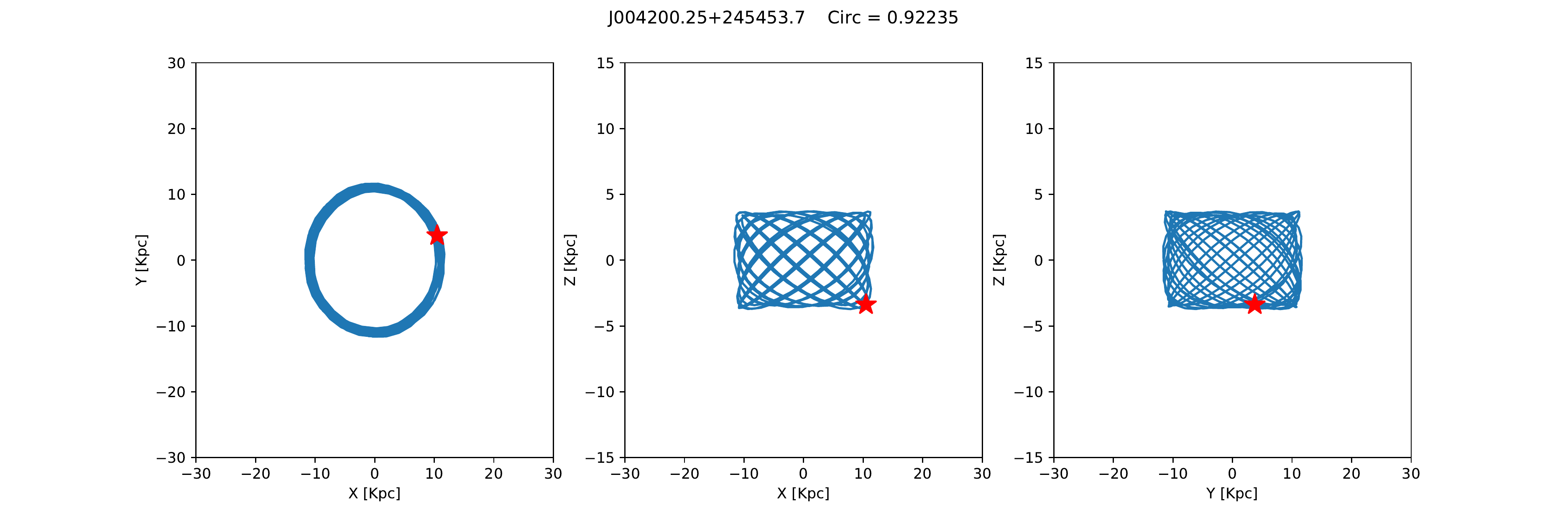}} \quad
            \subfloat[][]
        {\includegraphics[width=\textwidth]{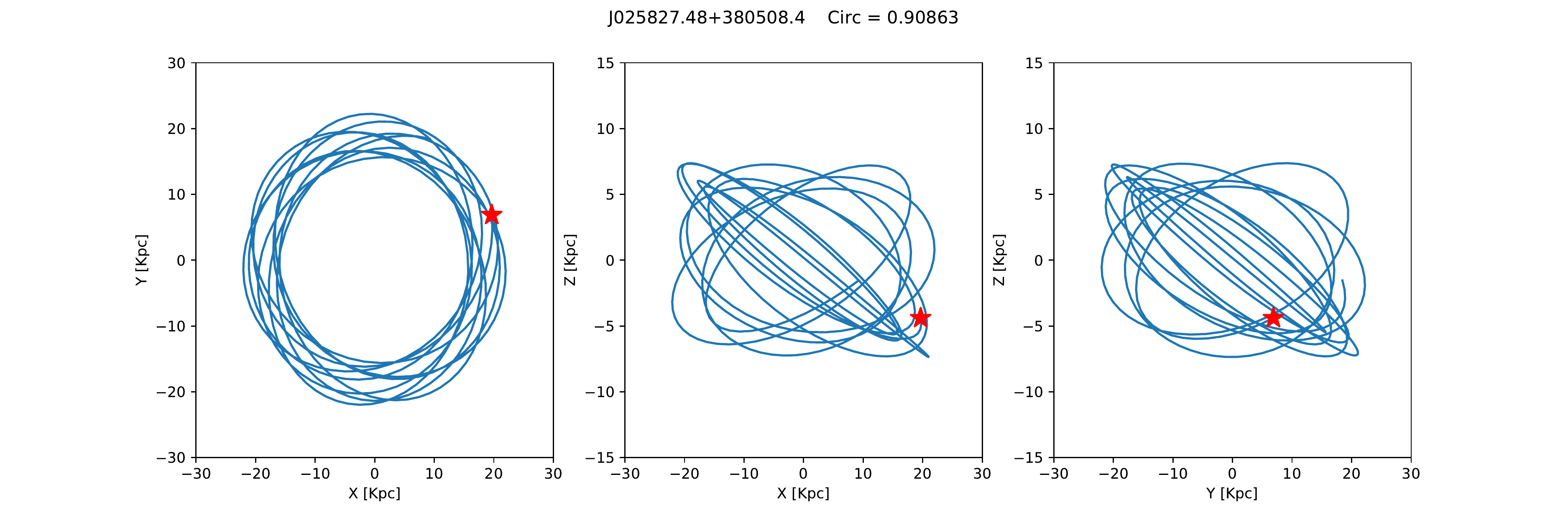}} \quad
            \subfloat[][]
        {\includegraphics[width=\textwidth]{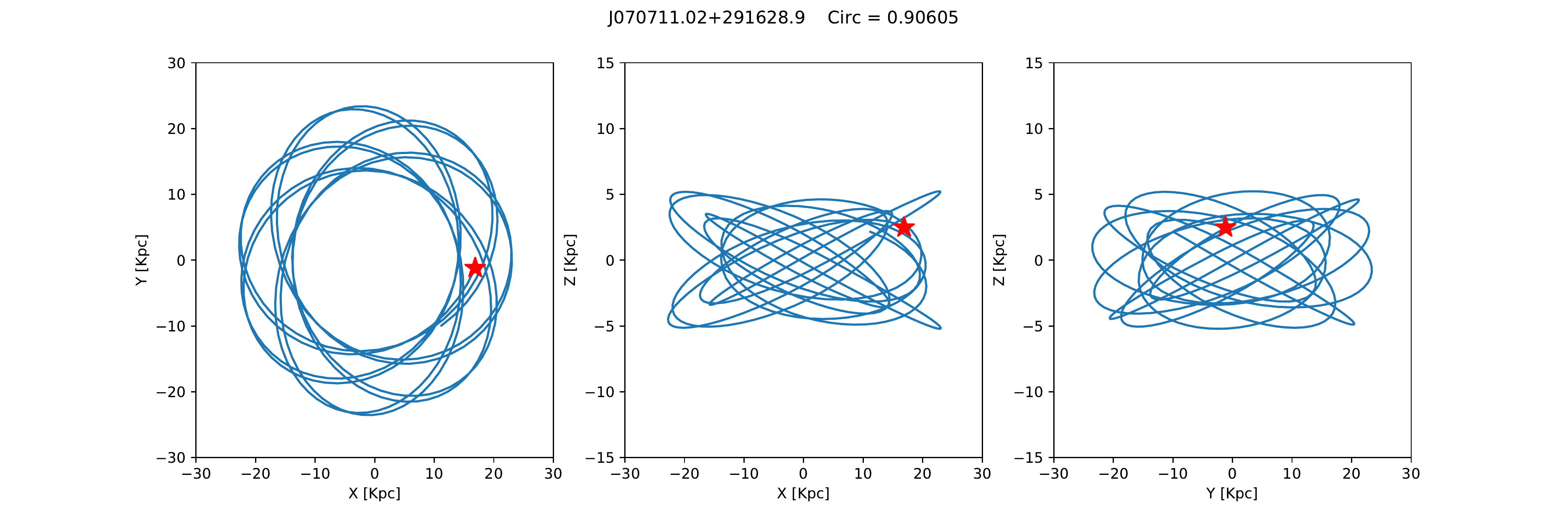}} \quad
        
    \caption{Continued.}
\end{figure*}

\begin{figure*}\ContinuedFloat
\captionsetup[subfigure]{labelformat=empty}
        \subfloat[][]
        {\includegraphics[width=\textwidth]{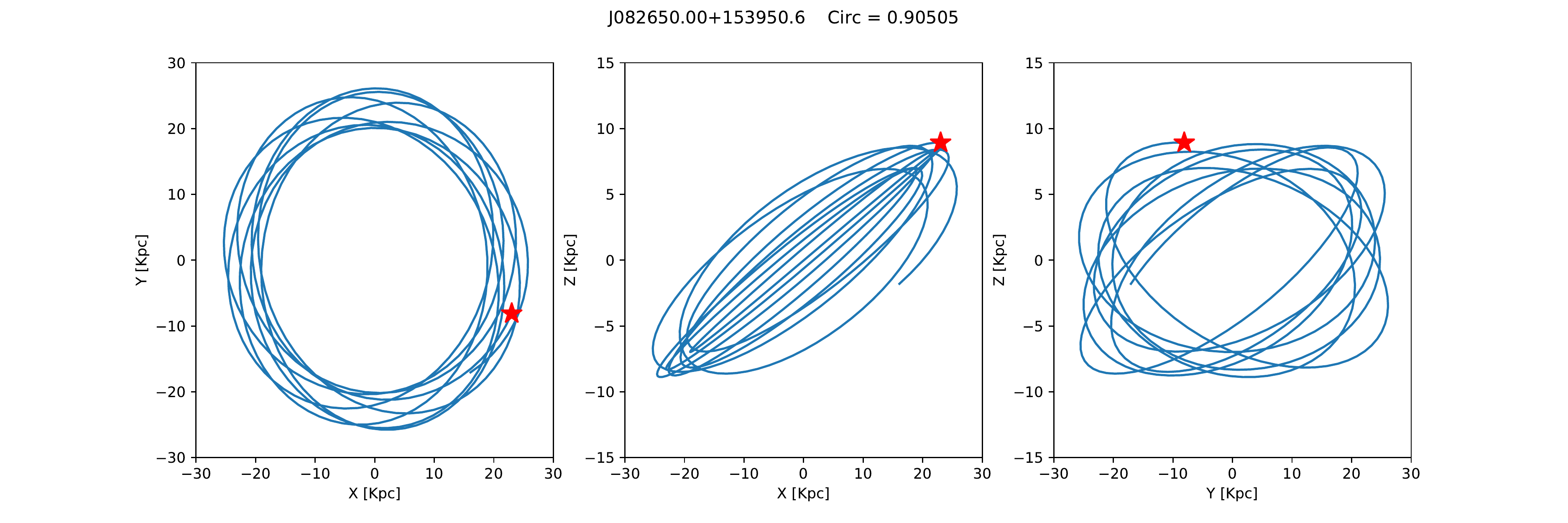}} \quad
            \subfloat[][]
        {\includegraphics[width=\textwidth]{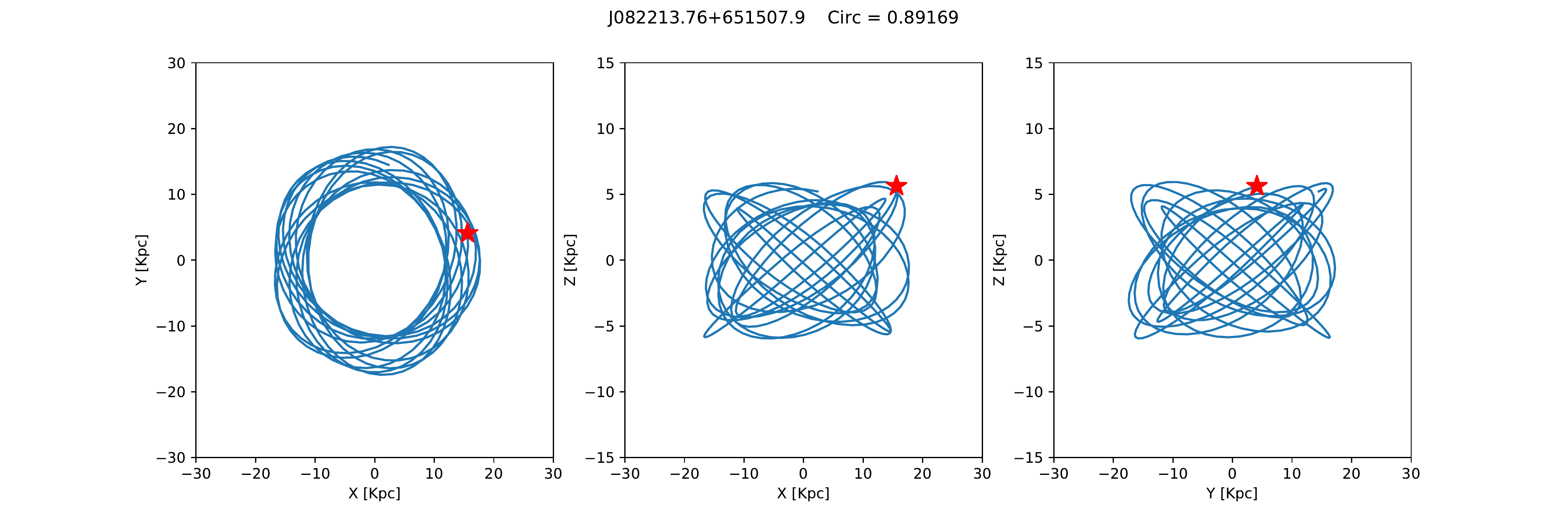}} \quad
            \subfloat[][]
        {\includegraphics[width=\textwidth]{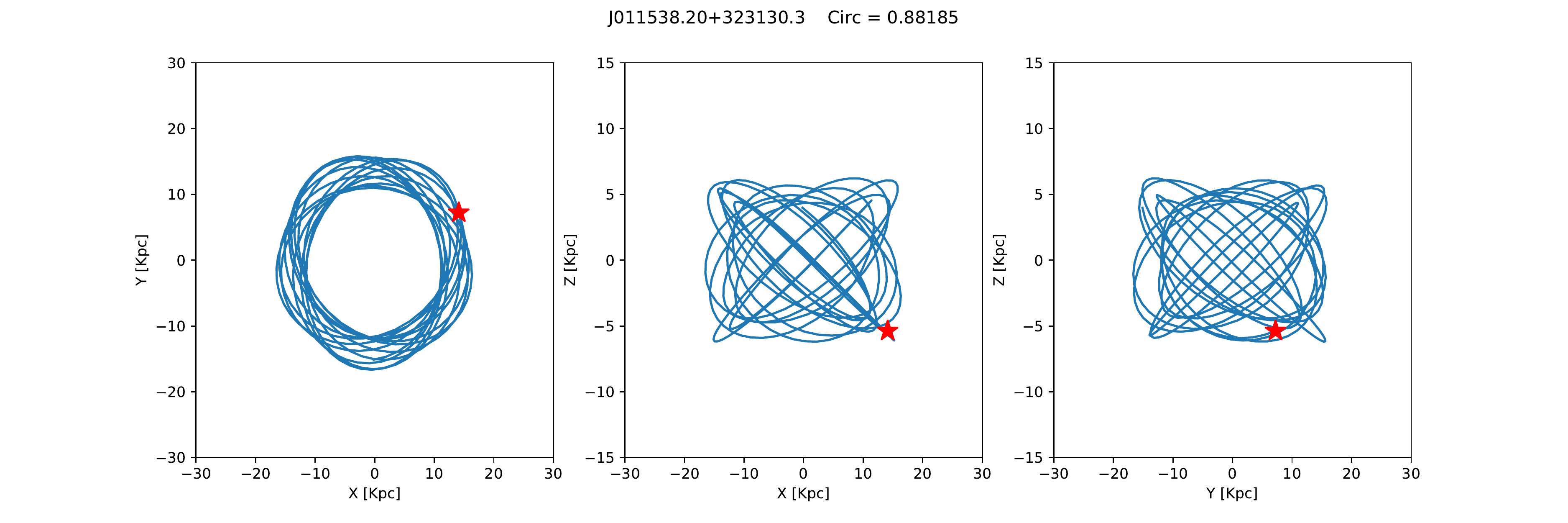}} \quad
        
    \caption{Continued.}
\end{figure*}

\begin{figure*}\ContinuedFloat
\captionsetup[subfigure]{labelformat=empty}
        \subfloat[][]
        {\includegraphics[width=\textwidth]{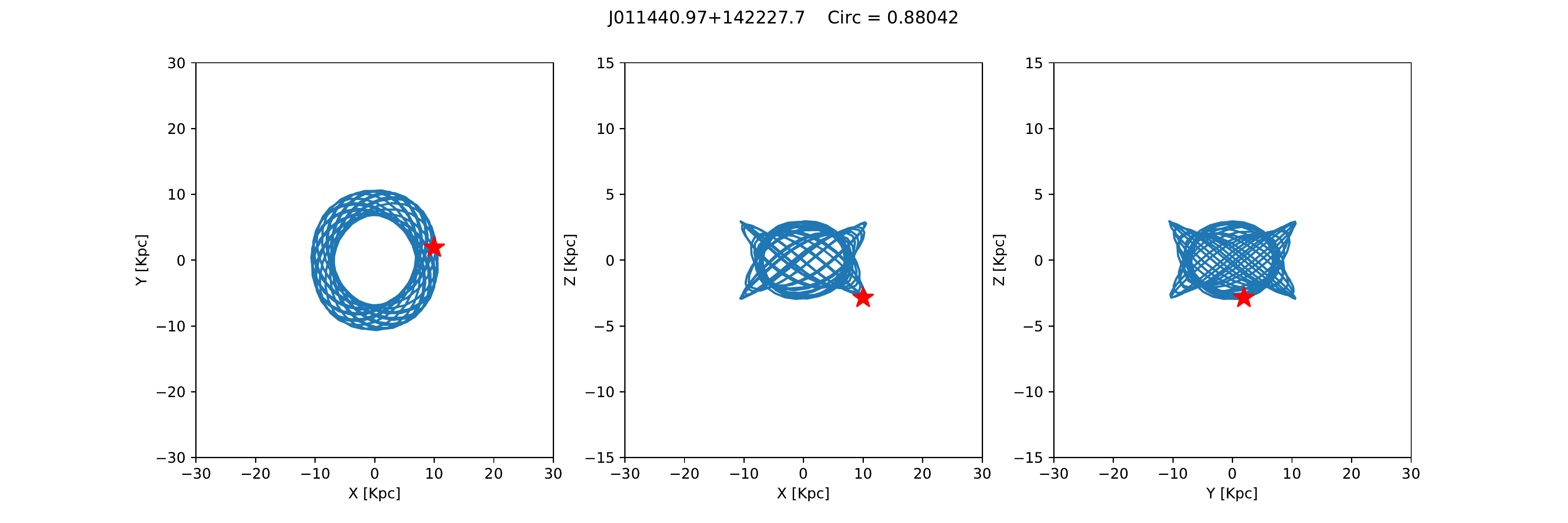}} \quad
            \subfloat[][]
        {\includegraphics[width=\textwidth]{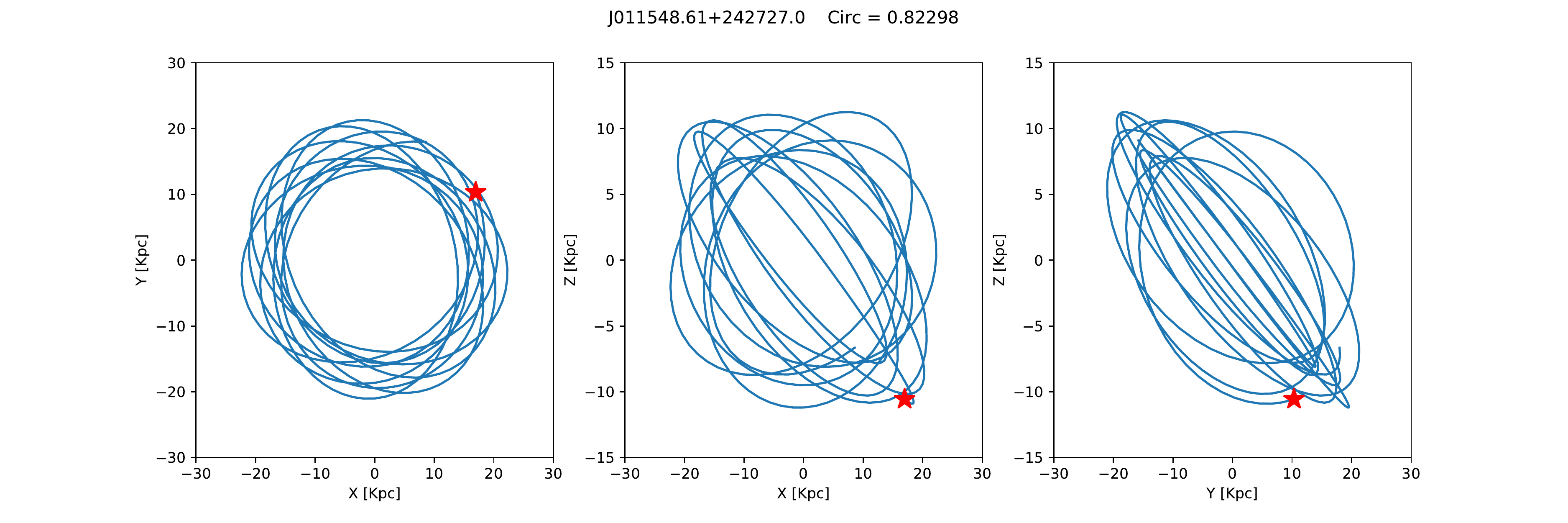}} \quad
            \subfloat[][]
        {\includegraphics[width=\textwidth]{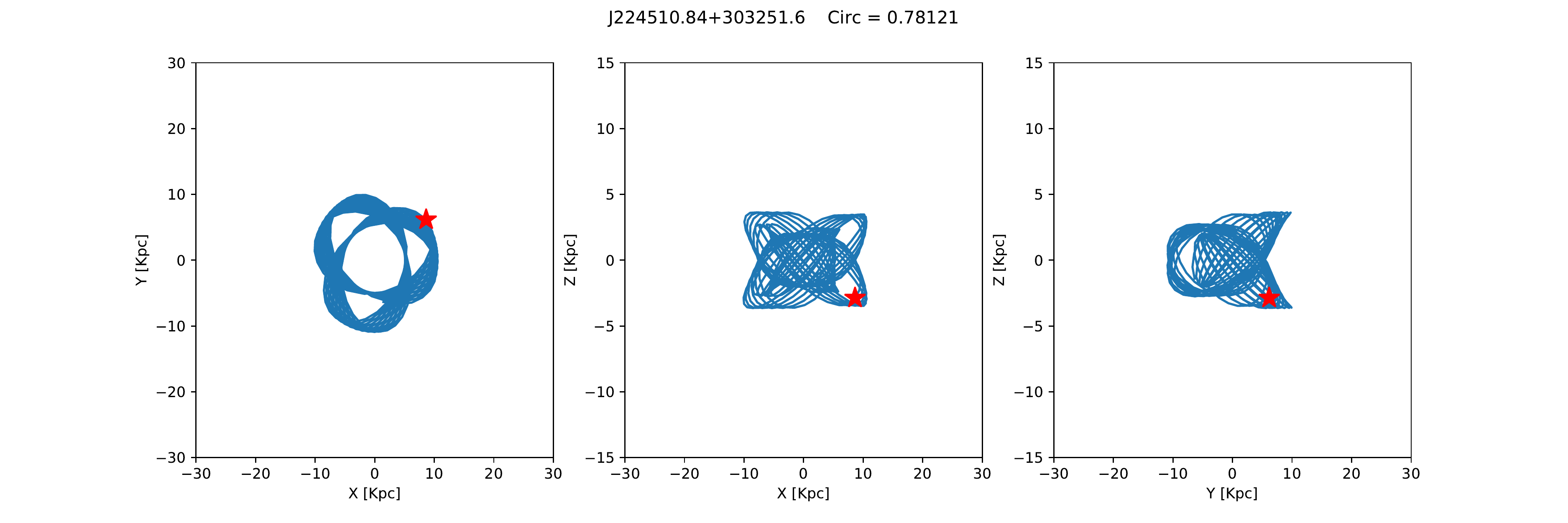}} \quad
        
    \caption{Continued.}
\end{figure*}

\begin{figure*}\ContinuedFloat
\captionsetup[subfigure]{labelformat=empty}
        \subfloat[][]
        {\includegraphics[width=\textwidth]{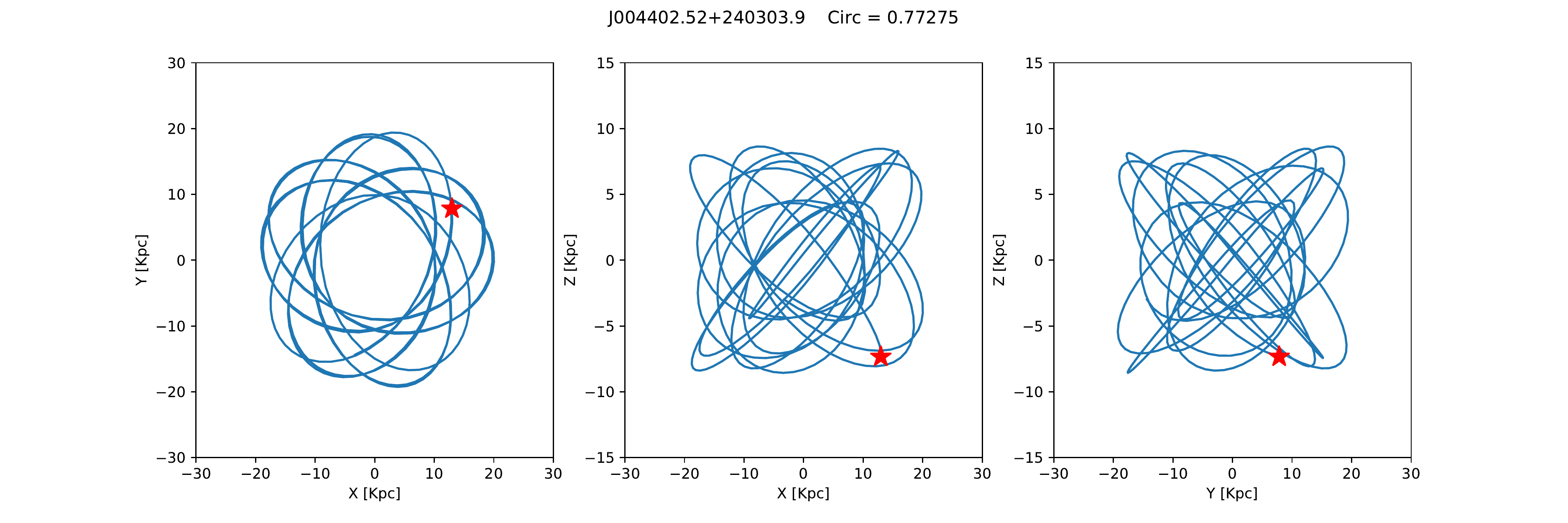}} \quad
            \subfloat[][]
        {\includegraphics[width=\textwidth]{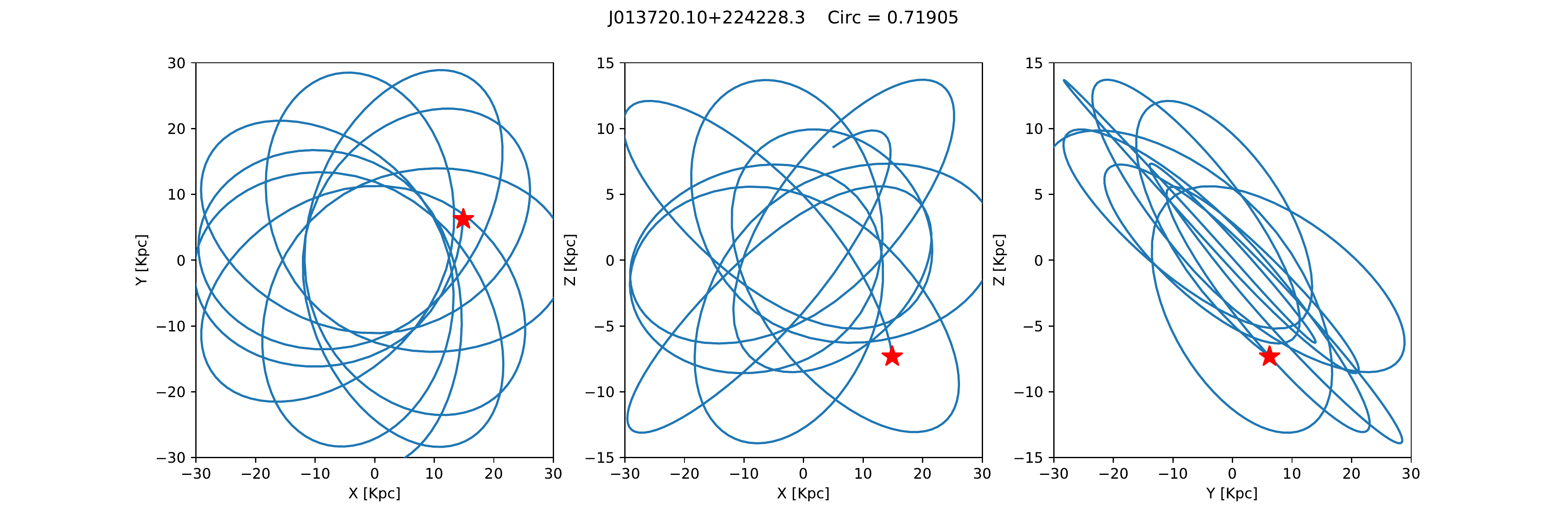}} \quad
            \subfloat[][]
        {\includegraphics[width=\textwidth]{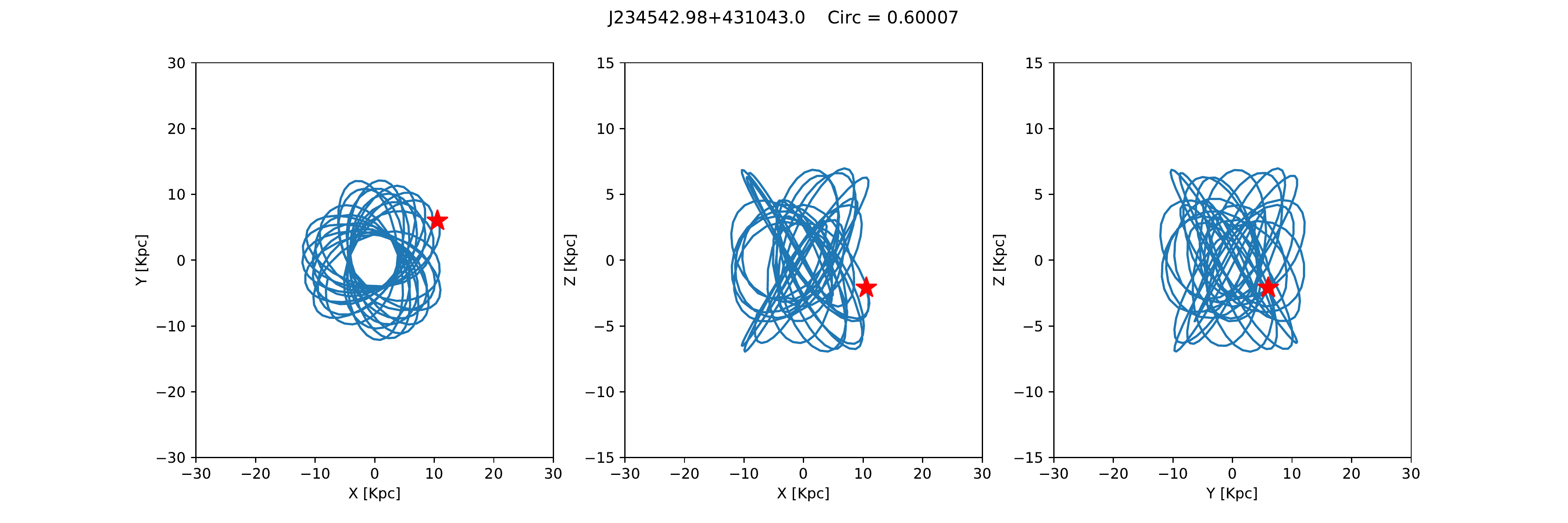}} \quad
        
    \caption{Continued.}
\end{figure*}

\begin{figure*}\ContinuedFloat
\captionsetup[subfigure]{labelformat=empty}
        \subfloat[][]
        {\includegraphics[width=\textwidth]{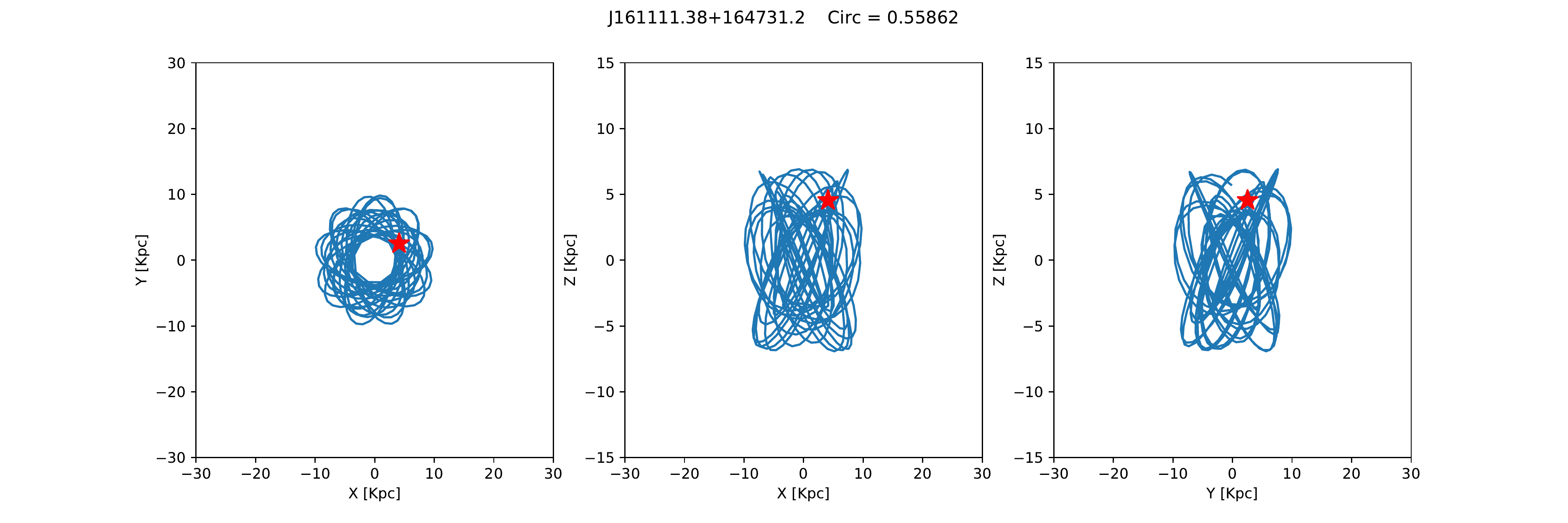}} \quad
            \subfloat[][]
        {\includegraphics[width=\textwidth]{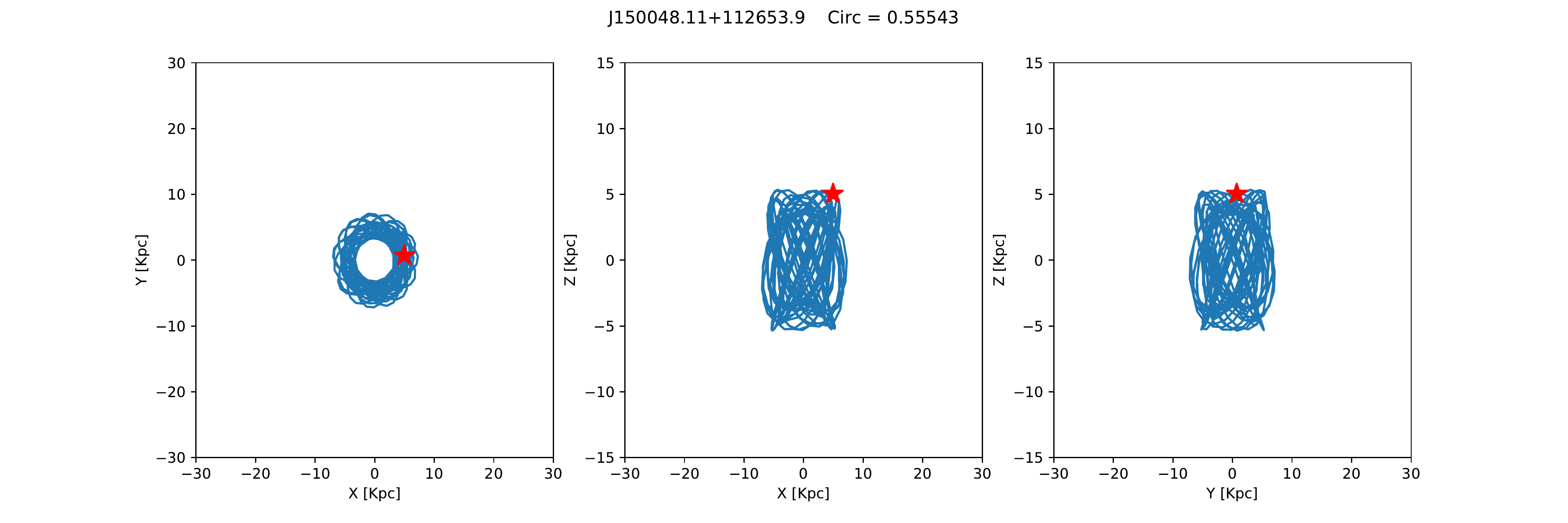}} \quad
            \subfloat[][]
        {\includegraphics[width=\textwidth]{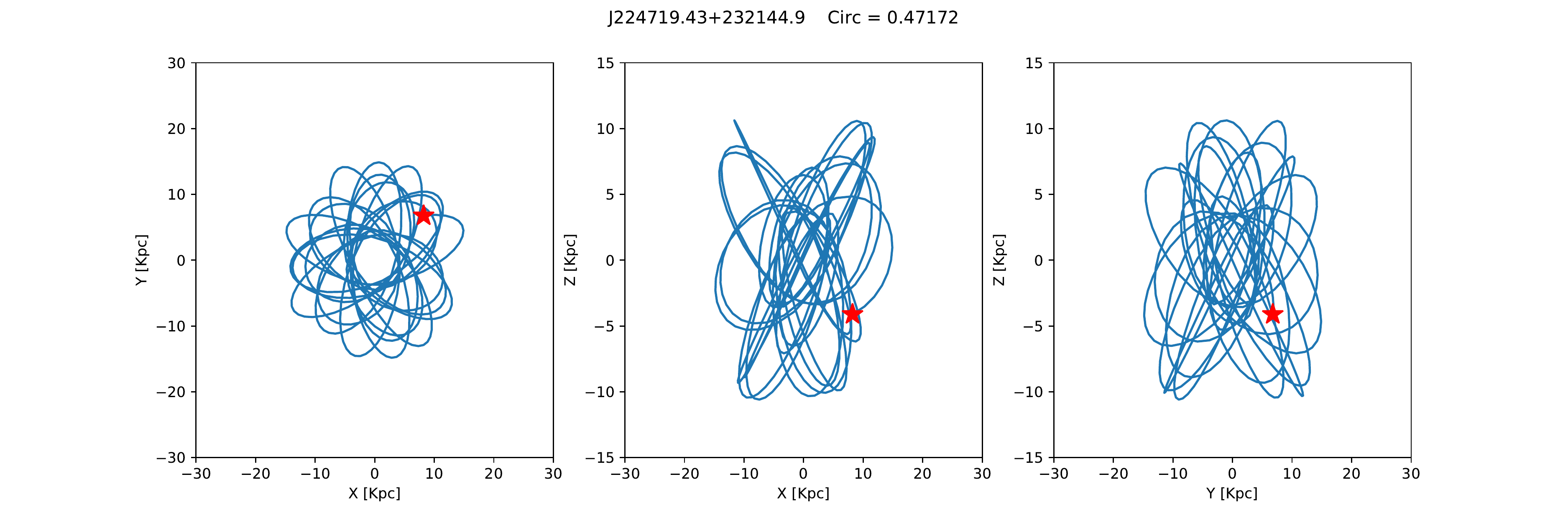}} \quad
        
    \caption{Continued.}
\end{figure*}

\begin{figure*}\ContinuedFloat
\captionsetup[subfigure]{labelformat=empty}
        \subfloat[][]
        {\includegraphics[width=\textwidth]{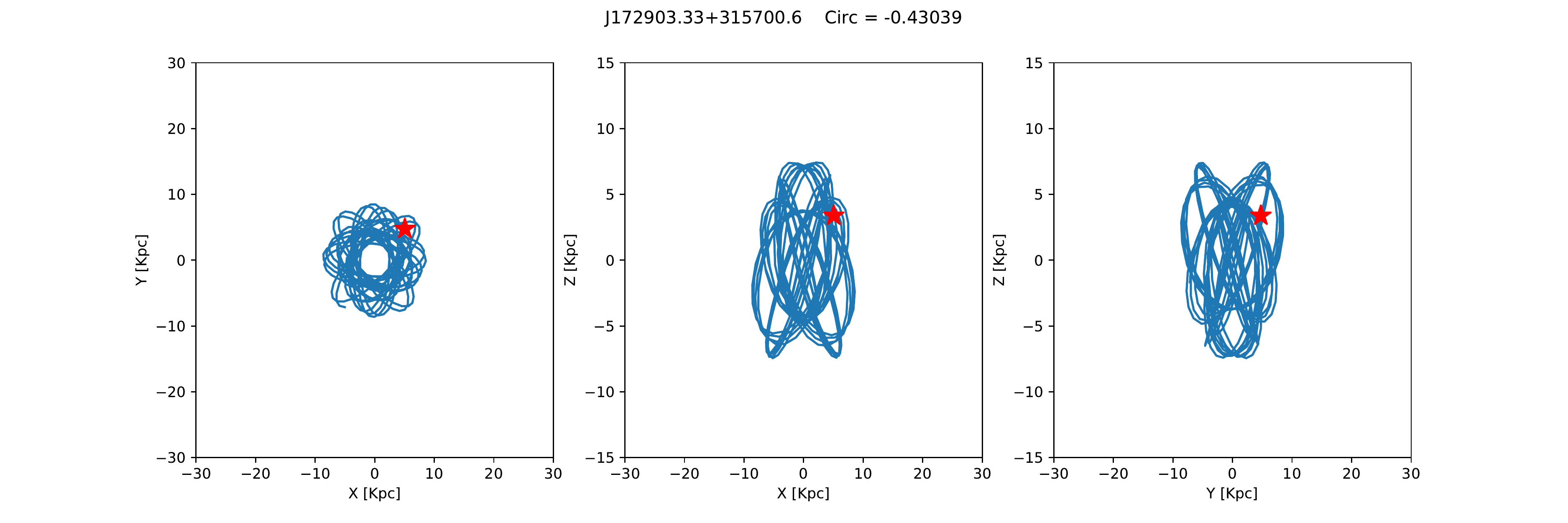}} \quad
            \subfloat[][]
        {\includegraphics[width=\textwidth]{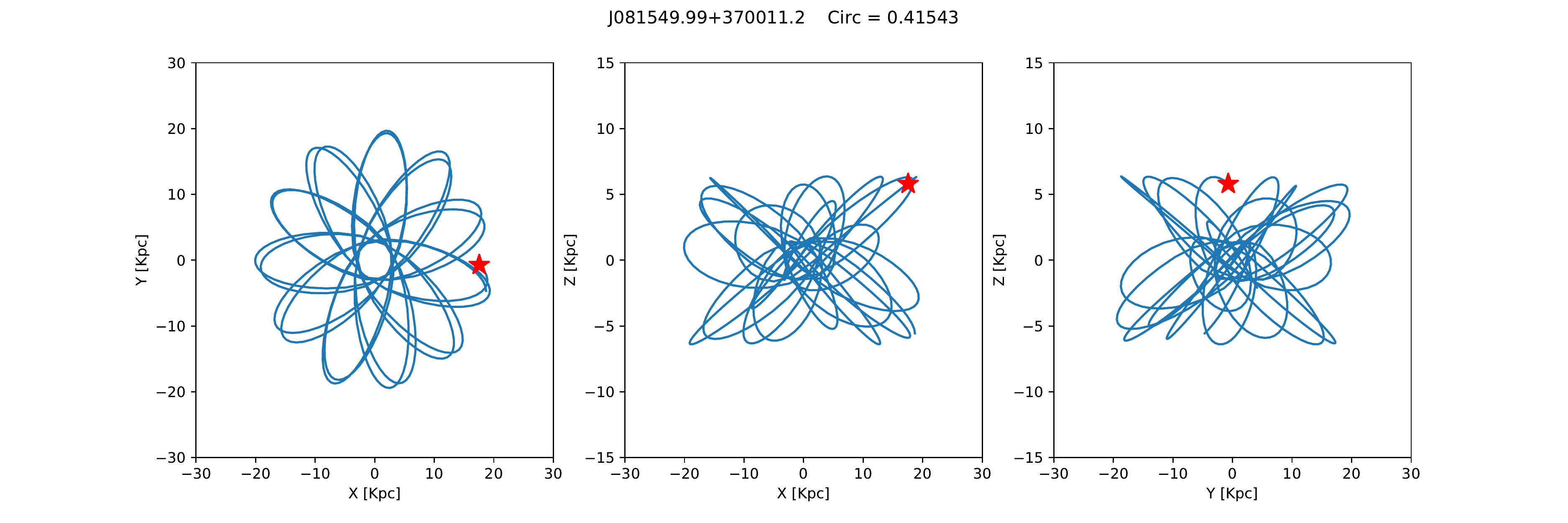}} \quad
            \subfloat[][]
        {\includegraphics[width=\textwidth]{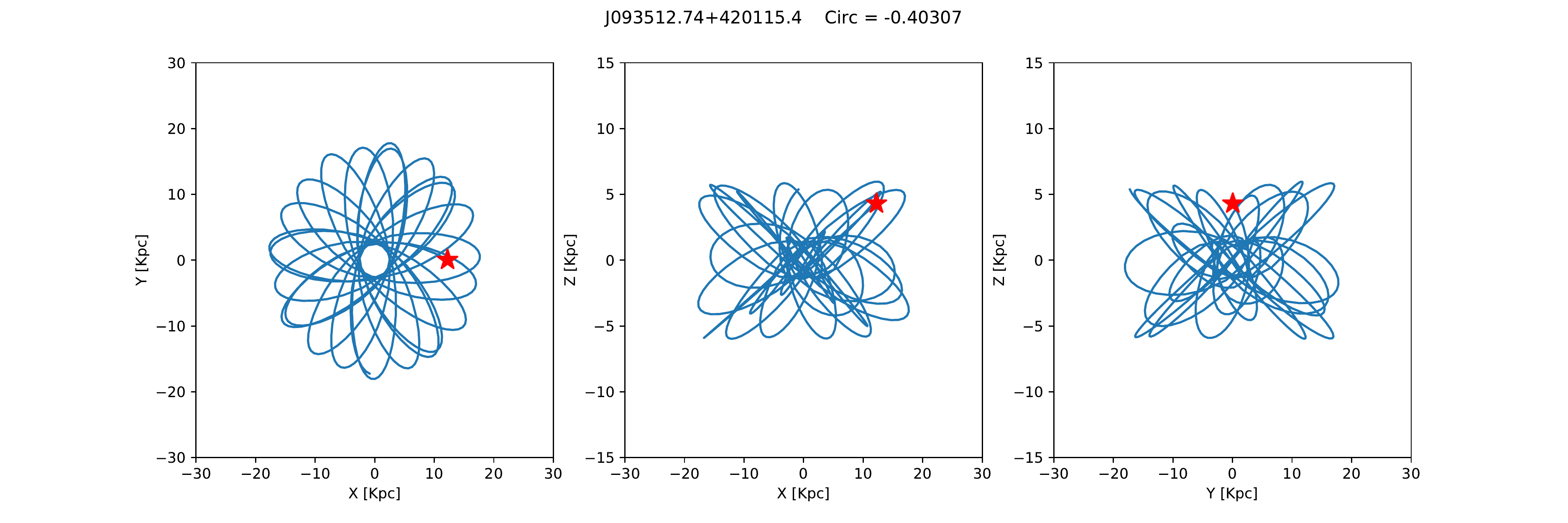}} \quad
        
    \caption{Continued.}
\end{figure*}

\begin{figure*}\ContinuedFloat
\captionsetup[subfigure]{labelformat=empty}
        \subfloat[][]
        {\includegraphics[width=\textwidth]{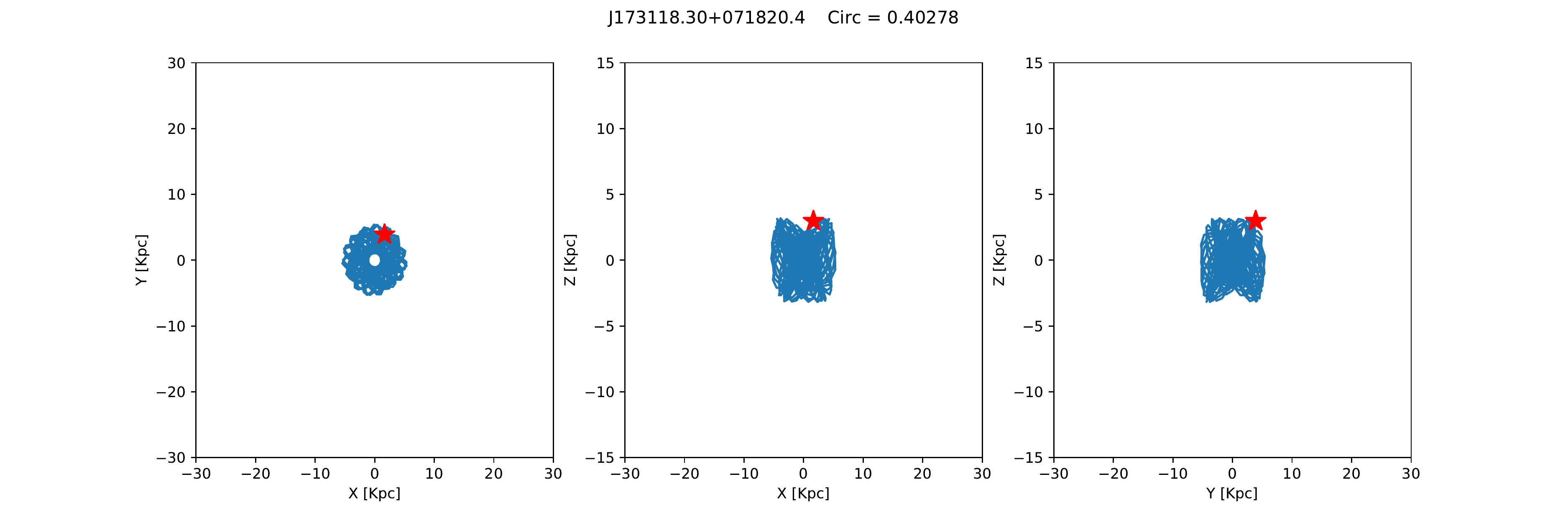}} \quad
            \subfloat[][]
        {\includegraphics[width=\textwidth]{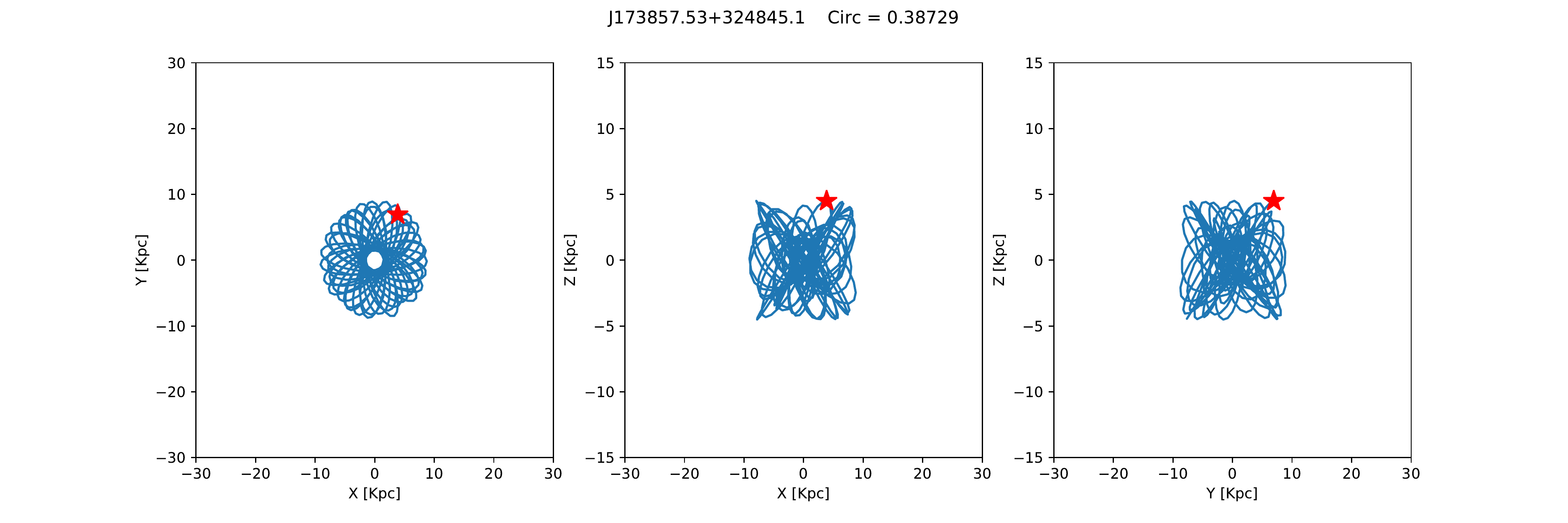}} \quad
            \subfloat[][]
        {\includegraphics[width=\textwidth]{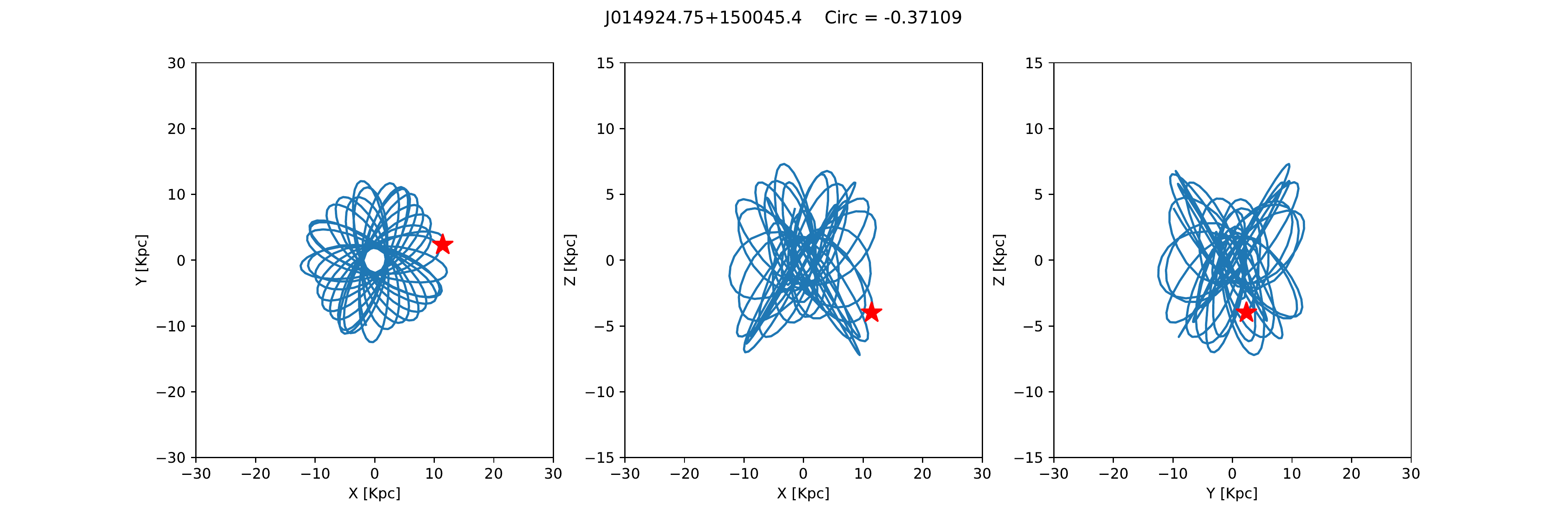}} \quad
        
    \caption{Continued.}
\end{figure*}

\begin{figure*}\ContinuedFloat
\captionsetup[subfigure]{labelformat=empty}
        \subfloat[][]
        {\includegraphics[width=\textwidth]{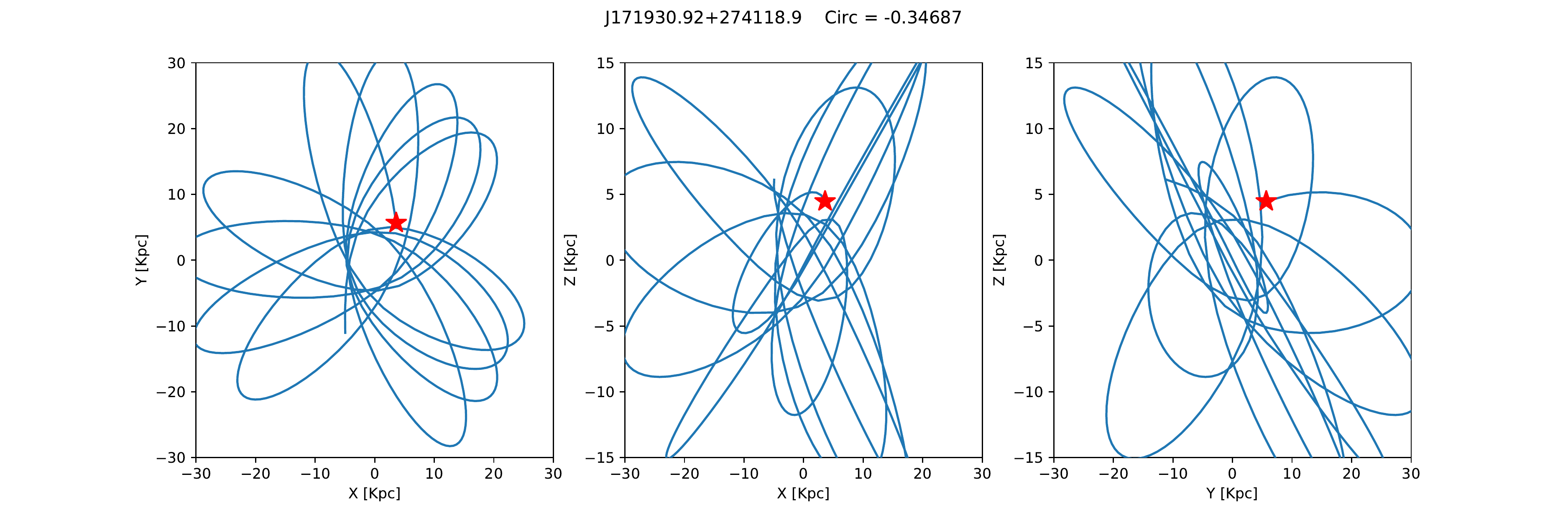}} \quad
            \subfloat[][]
        {\includegraphics[width=\textwidth]{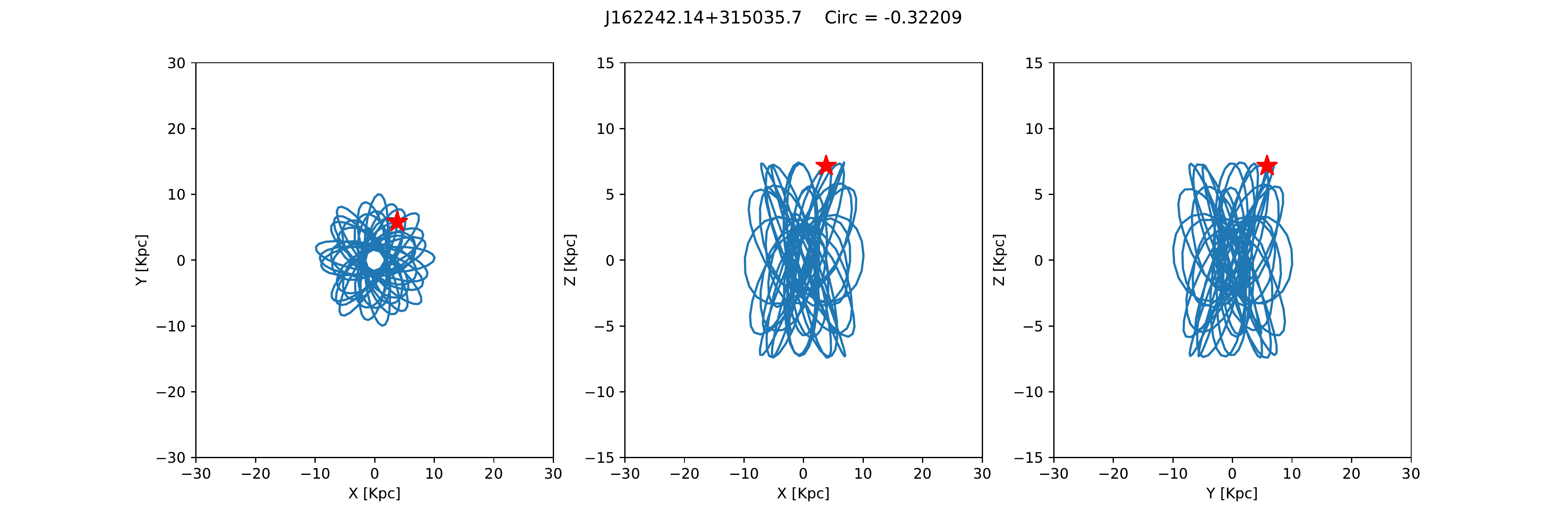}} \quad
            \subfloat[][]
        {\includegraphics[width=\textwidth]{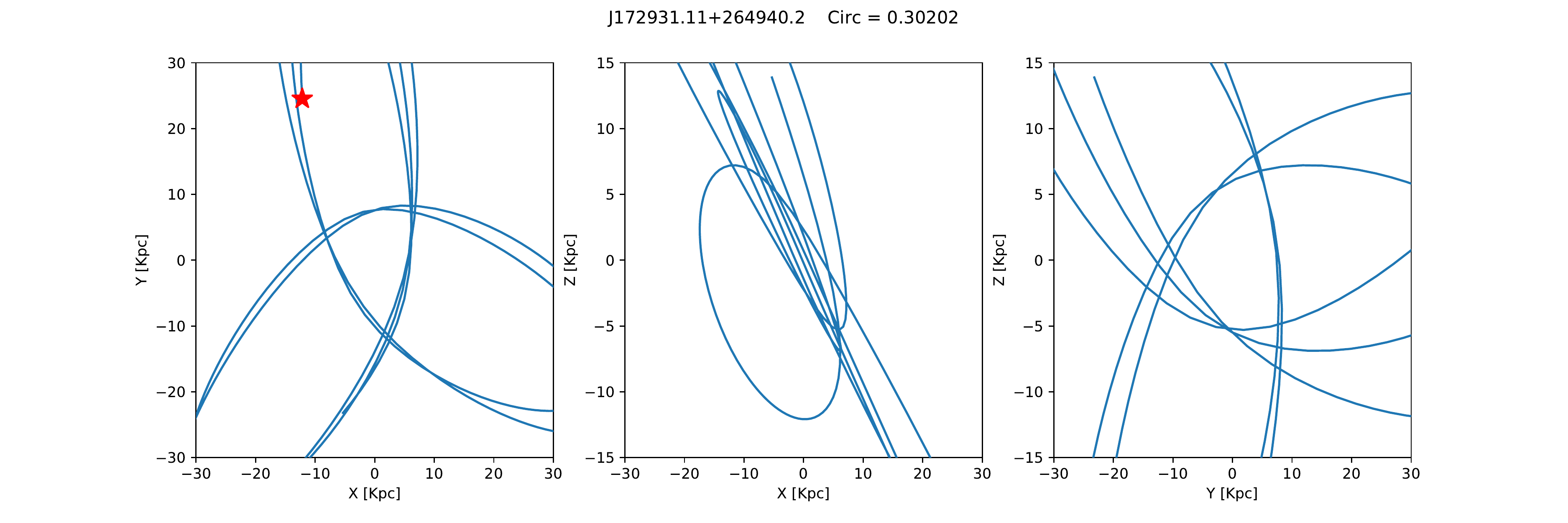}} \quad
        
    \caption{Continued.}
\end{figure*}

\begin{figure*}\ContinuedFloat
\captionsetup[subfigure]{labelformat=empty}
        \subfloat[][]
        {\includegraphics[width=\textwidth]{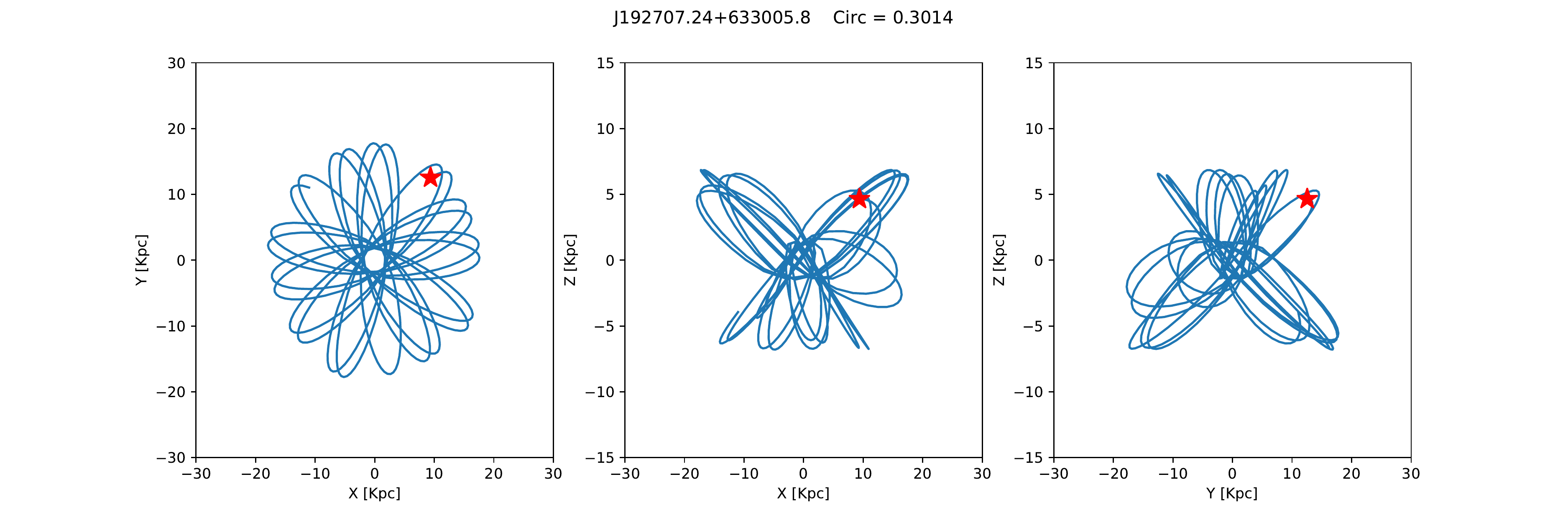}} \quad
            \subfloat[][]
        {\includegraphics[width=\textwidth]{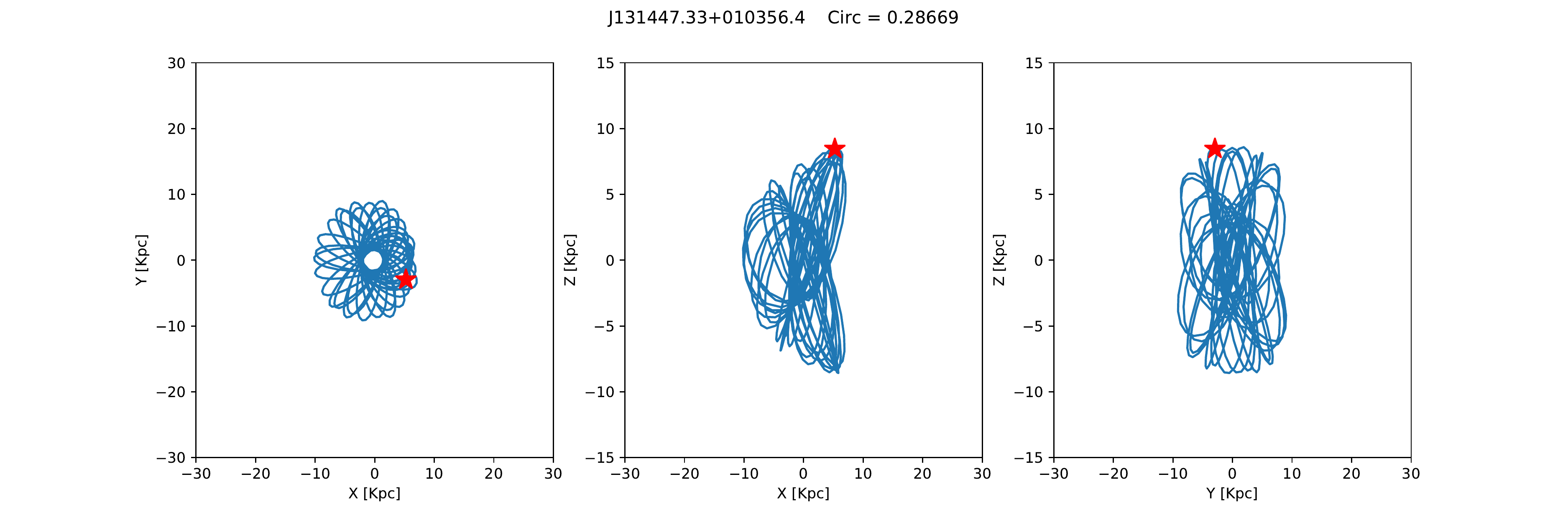}} \quad
            \subfloat[][]
        {\includegraphics[width=\textwidth]{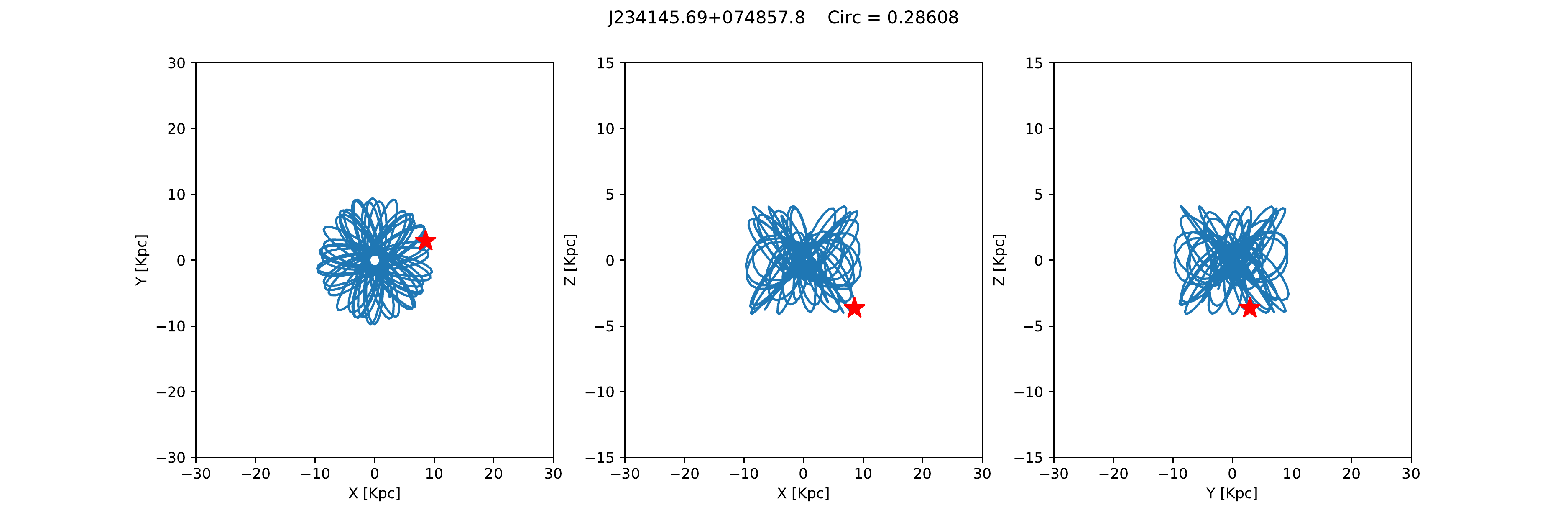}} \quad
        
    \caption{Continued.}
\end{figure*}

\begin{figure*}\ContinuedFloat
\captionsetup[subfigure]{labelformat=empty}
        \subfloat[][]
        {\includegraphics[width=\textwidth]{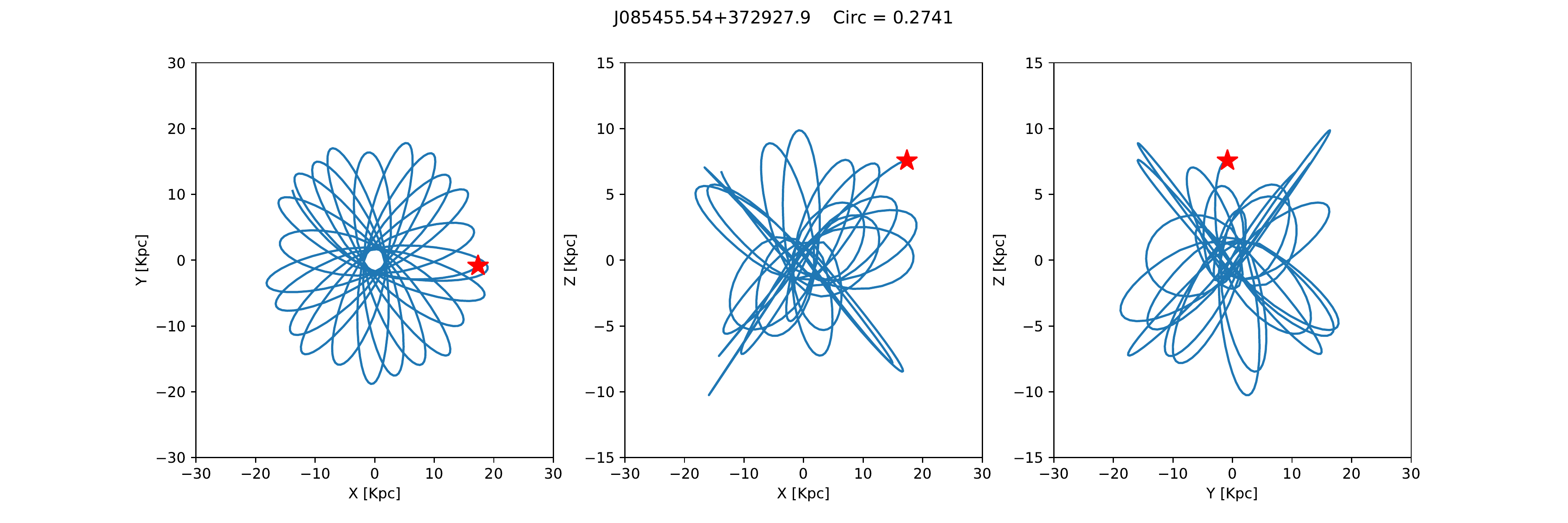}} \quad
            \subfloat[][]
        {\includegraphics[width=\textwidth]{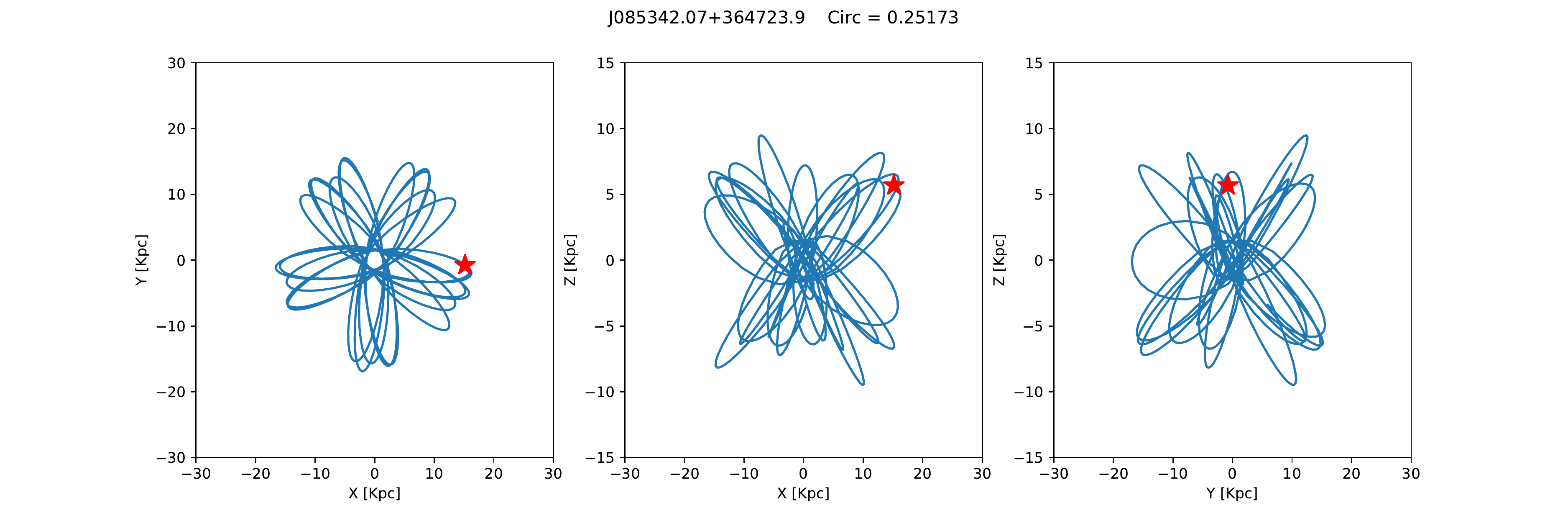}} \quad
            \subfloat[][]
        {\includegraphics[width=\textwidth]{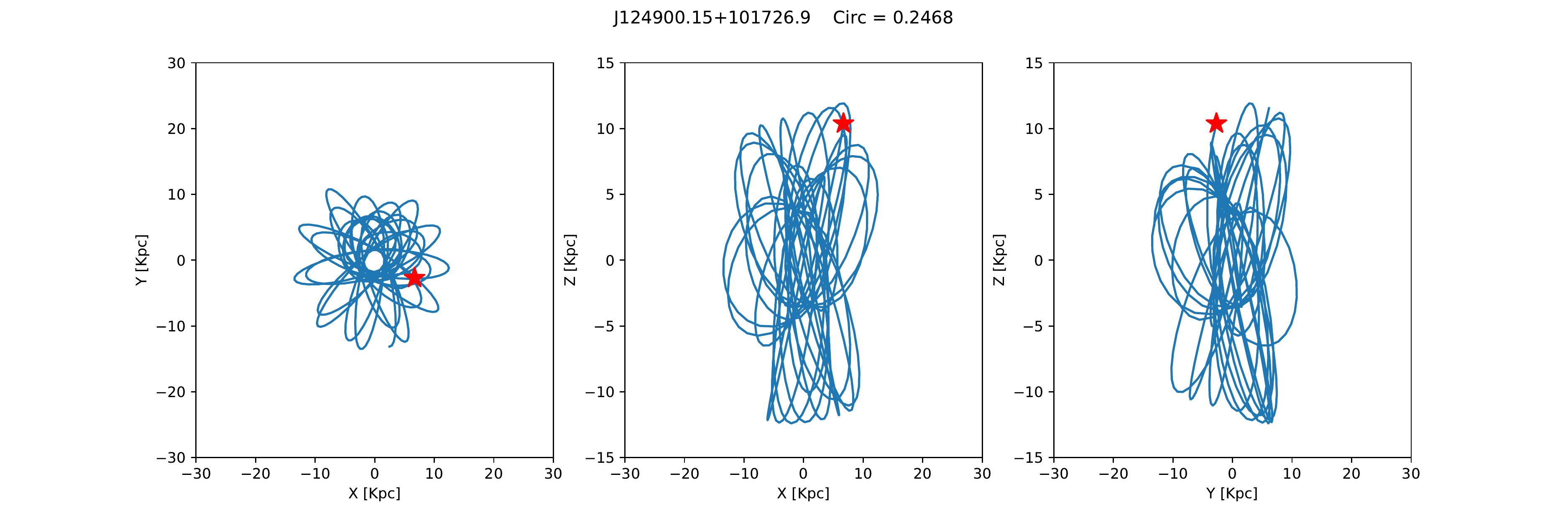}} \quad
        
    \caption{Continued.}
\end{figure*}

\begin{figure*}\ContinuedFloat
\captionsetup[subfigure]{labelformat=empty}
        \subfloat[][]
        {\includegraphics[width=\textwidth]{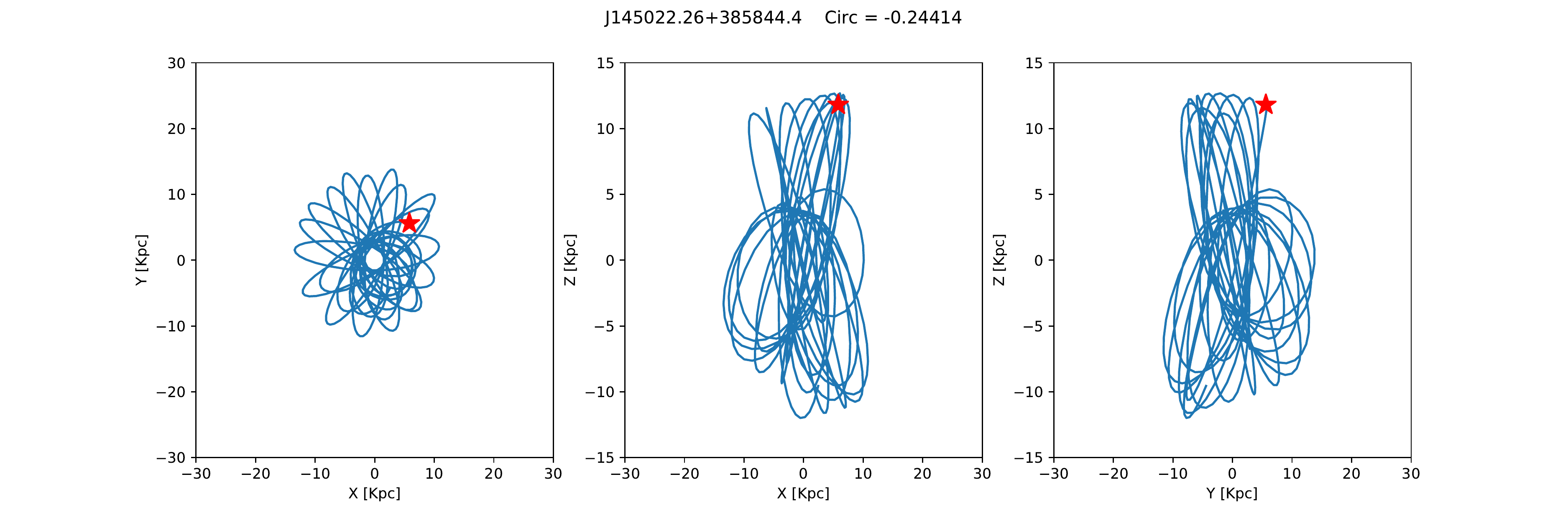}} \quad
            \subfloat[][]
        {\includegraphics[width=\textwidth]{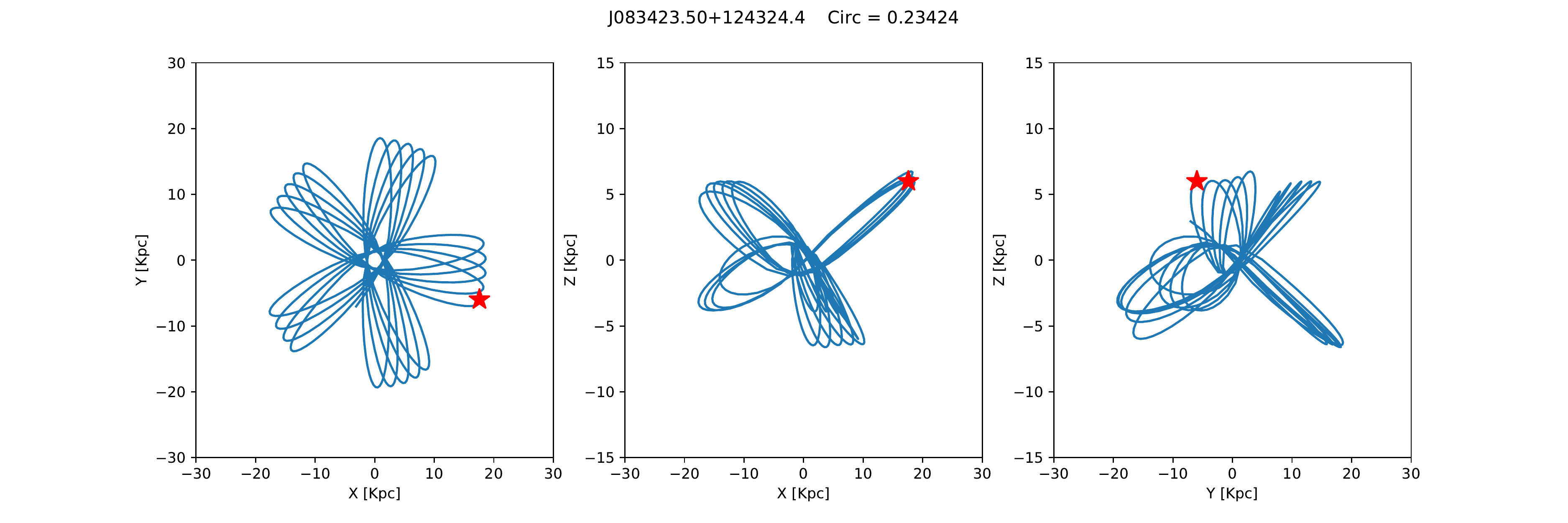}} \quad
            \subfloat[][]
        {\includegraphics[width=\textwidth]{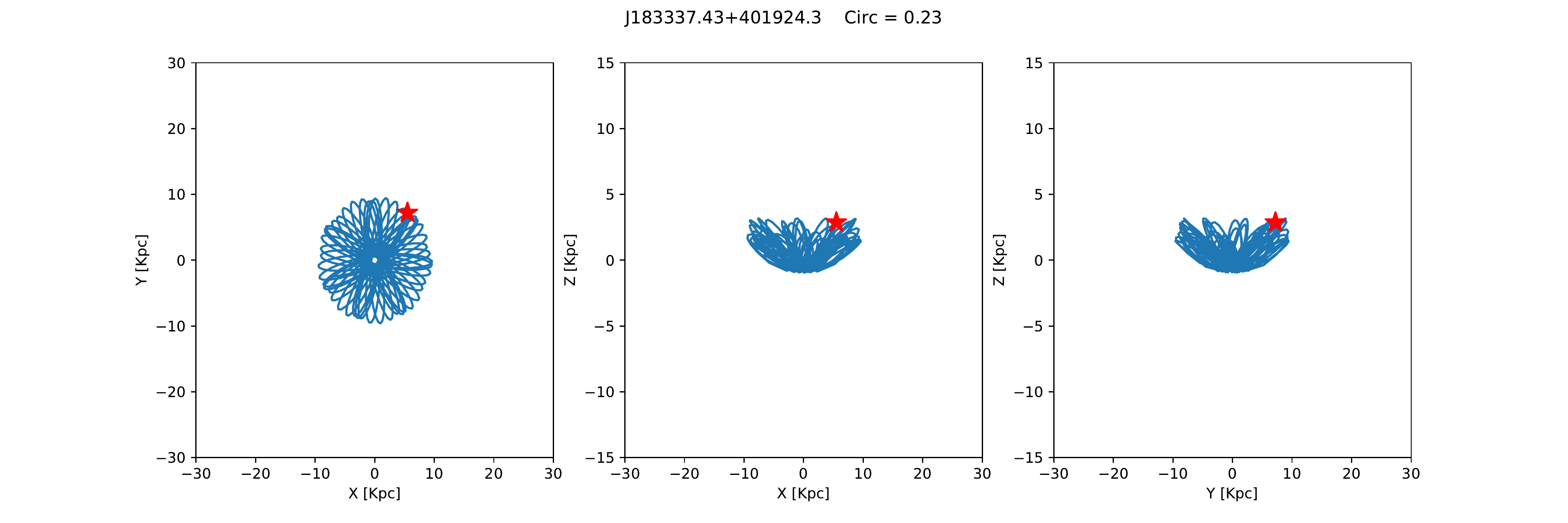}} \quad
        
    \caption{Continued.}
\end{figure*}

\begin{figure*}\ContinuedFloat
\captionsetup[subfigure]{labelformat=empty}
        \subfloat[][]
        {\includegraphics[width=\textwidth]{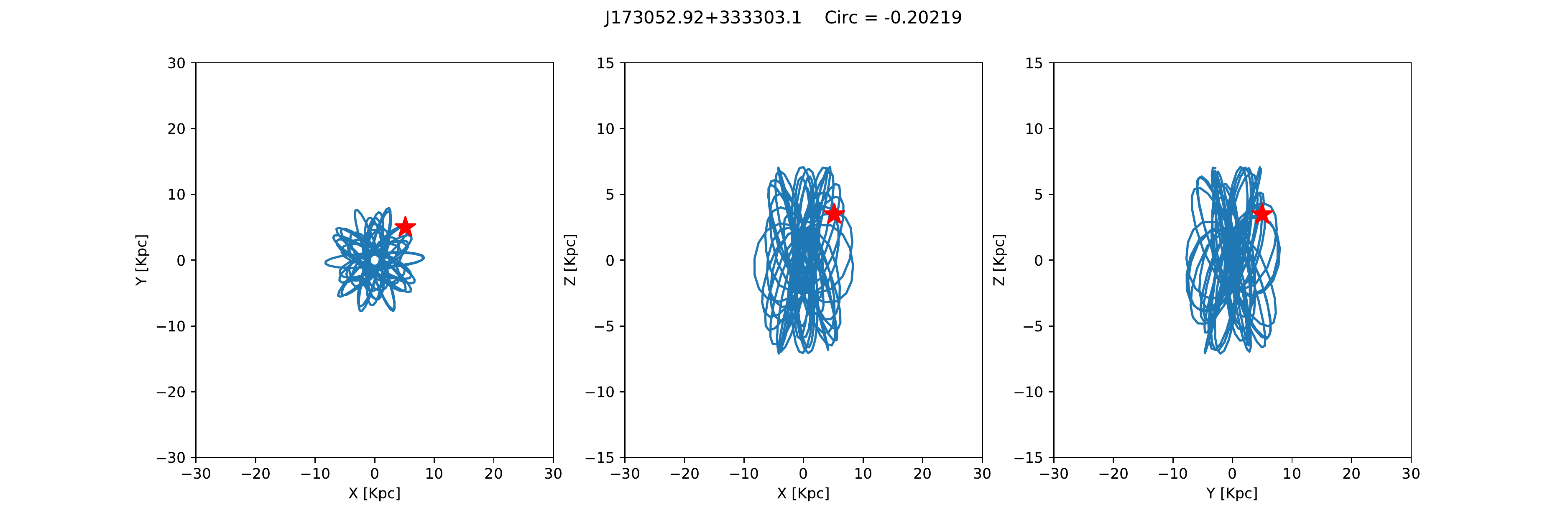}} \quad
            \subfloat[][]
        {\includegraphics[width=\textwidth]{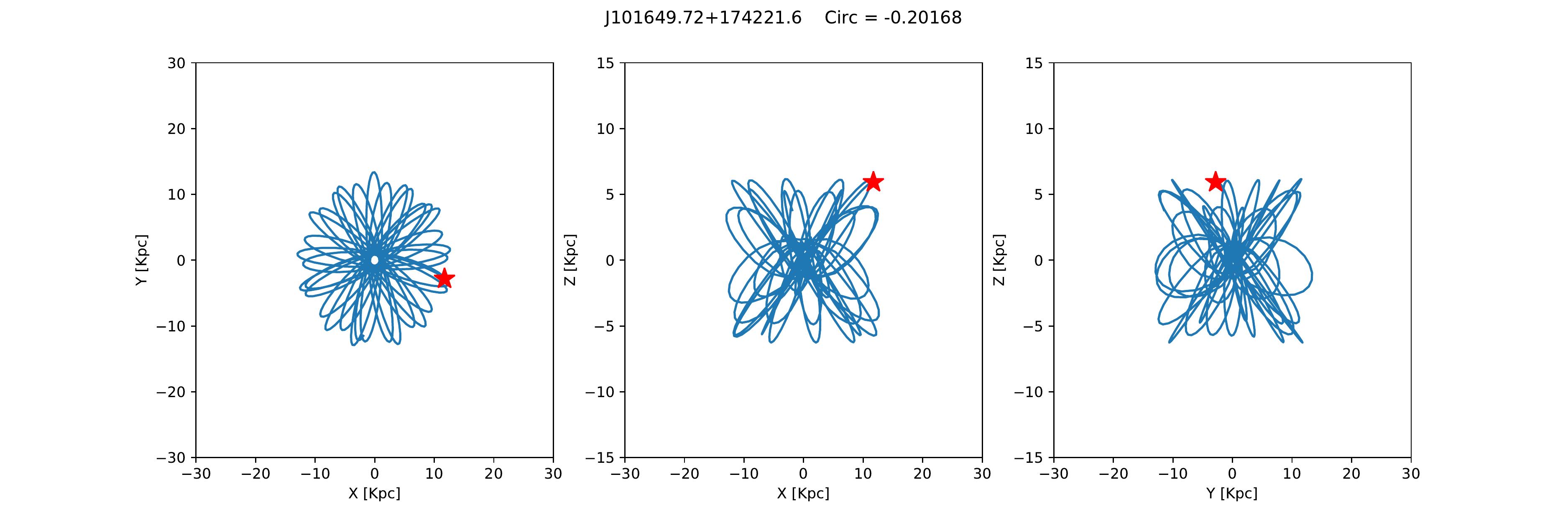}} \quad
            \subfloat[][]
        {\includegraphics[width=\textwidth]{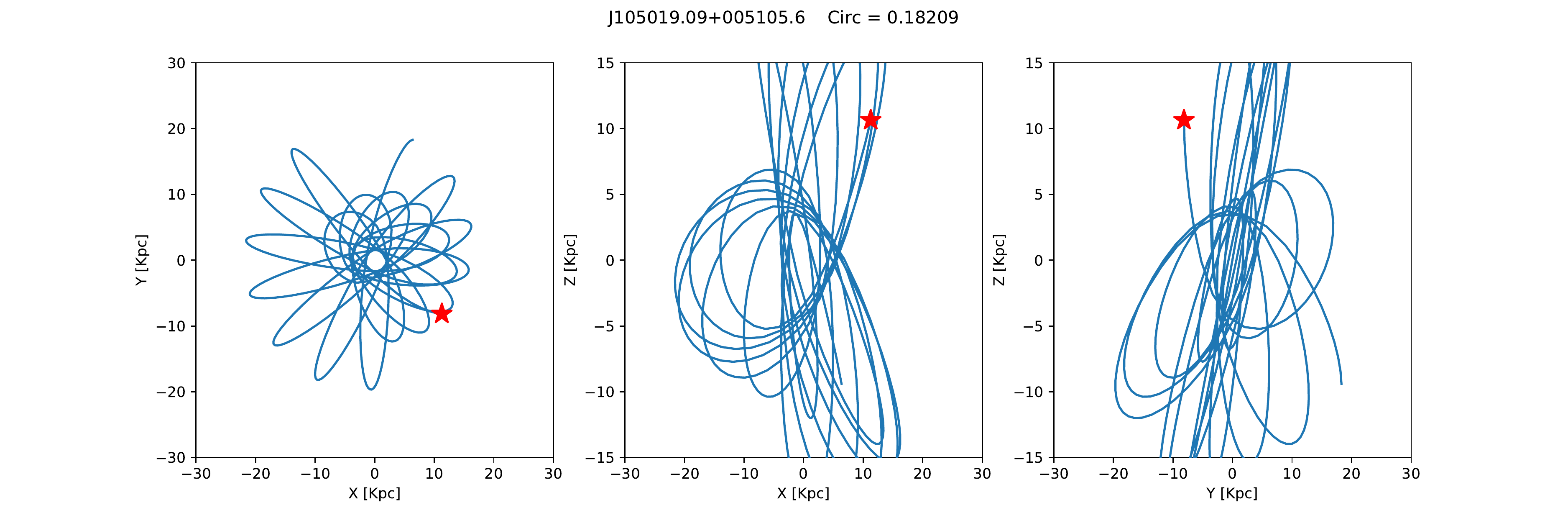}} \quad
        
    \caption{Continued.}
\end{figure*}

\begin{figure*}\ContinuedFloat
\captionsetup[subfigure]{labelformat=empty}
        \subfloat[][]
        {\includegraphics[width=\textwidth]{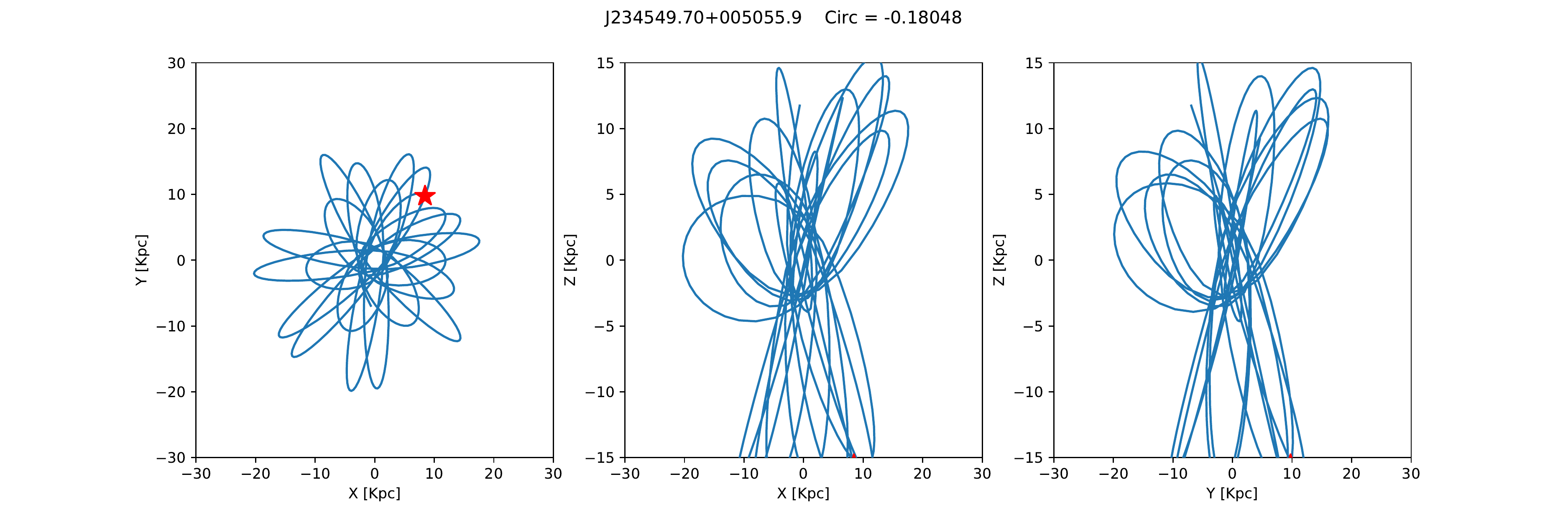}} \quad
            \subfloat[][]
        {\includegraphics[width=\textwidth]{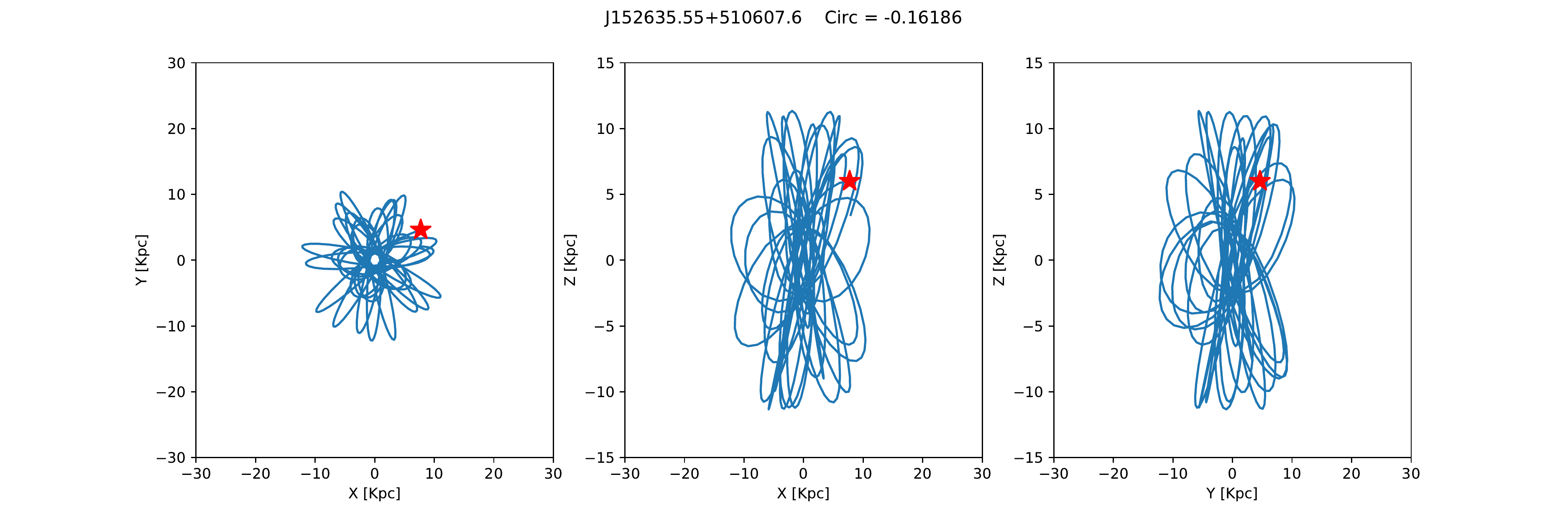}} \quad
            \subfloat[][]
        {\includegraphics[width=\textwidth]{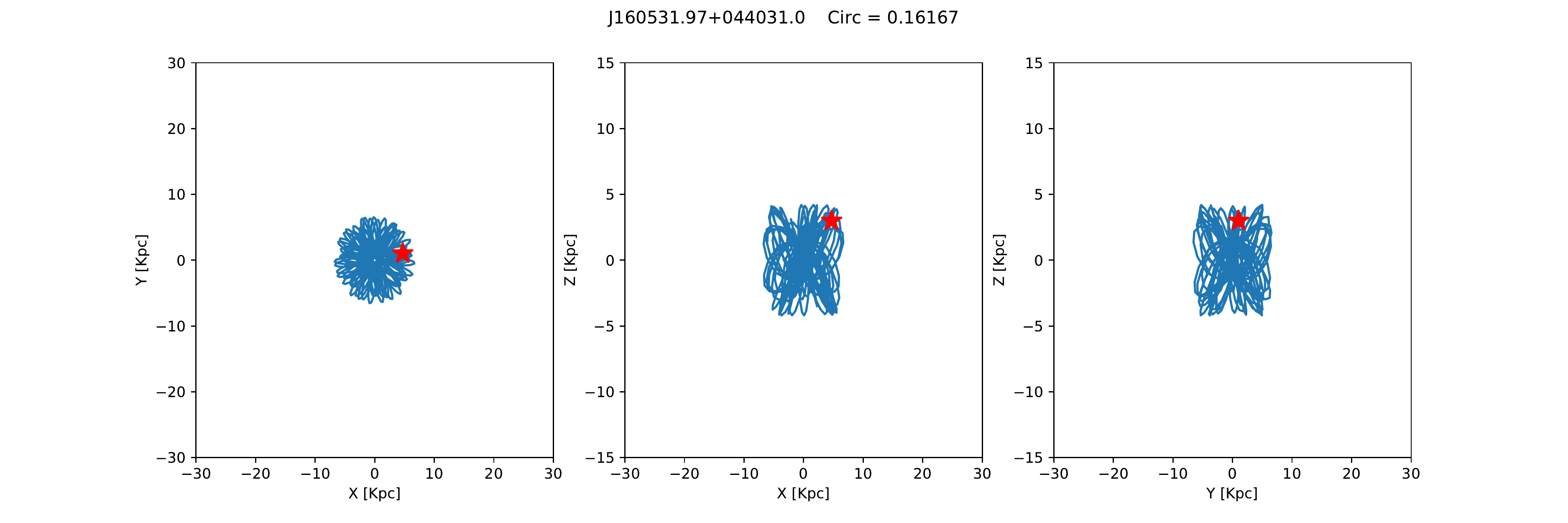}} \quad
        
    \caption{Continued.}
\end{figure*}

\begin{figure*}\ContinuedFloat
\captionsetup[subfigure]{labelformat=empty}
        \subfloat[][]
        {\includegraphics[width=\textwidth]{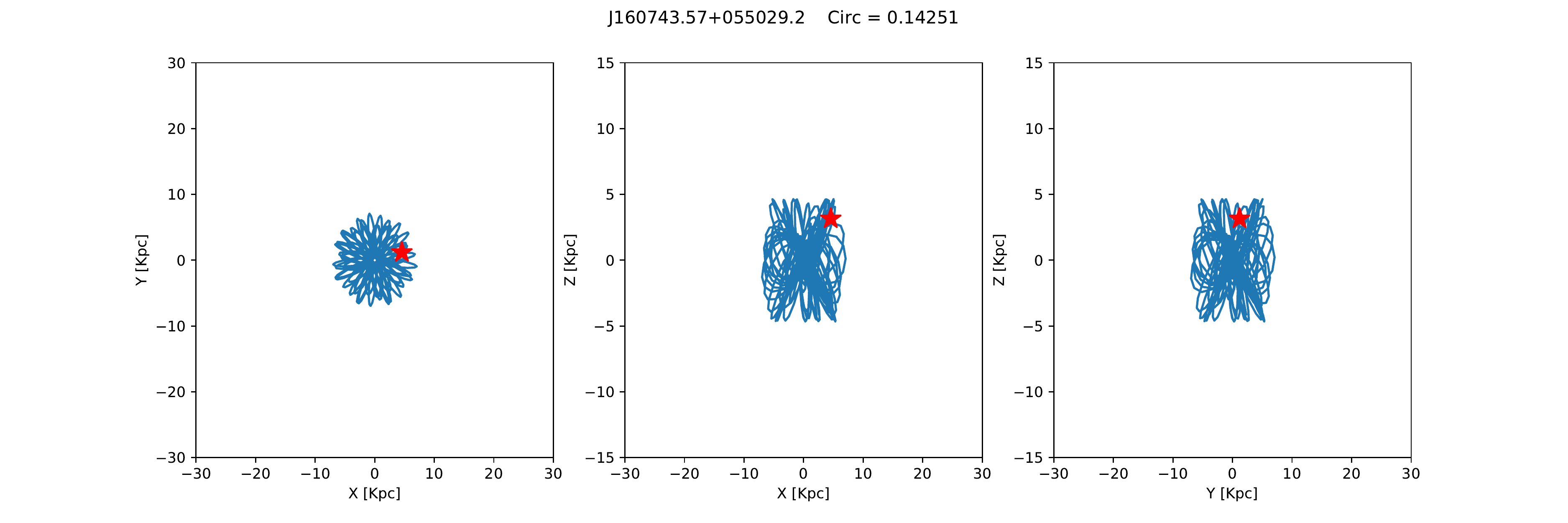}} \quad
            \subfloat[][]
        {\includegraphics[width=\textwidth]{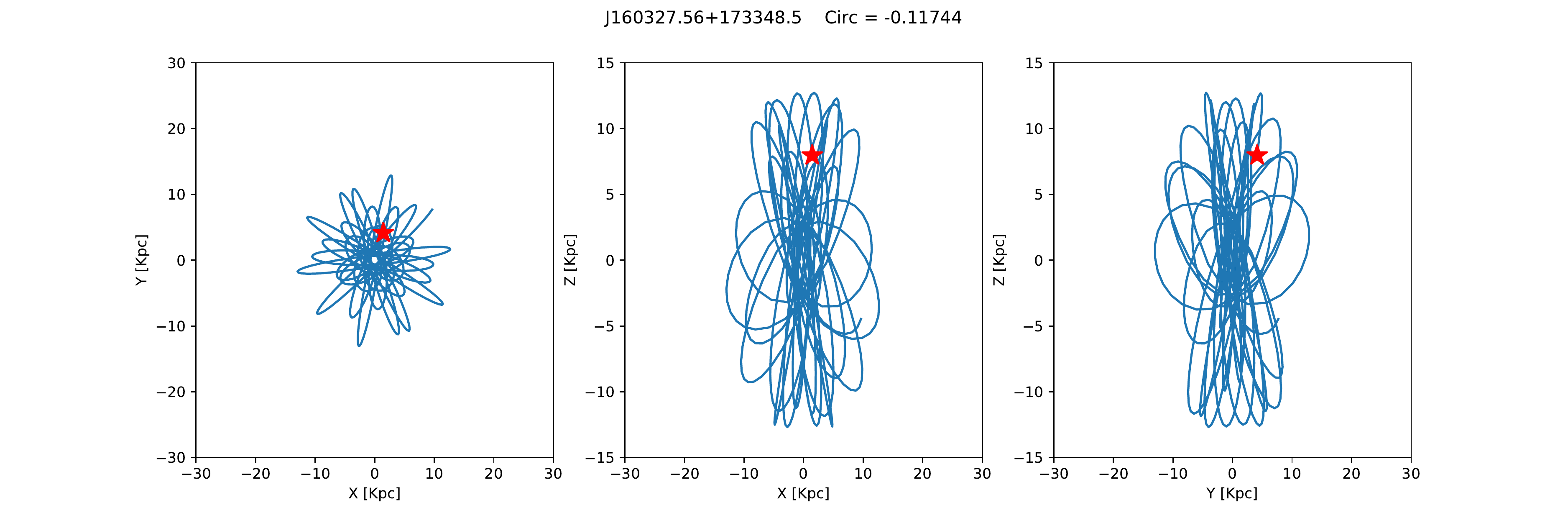}} \quad
            \subfloat[][]
        {\includegraphics[width=\textwidth]{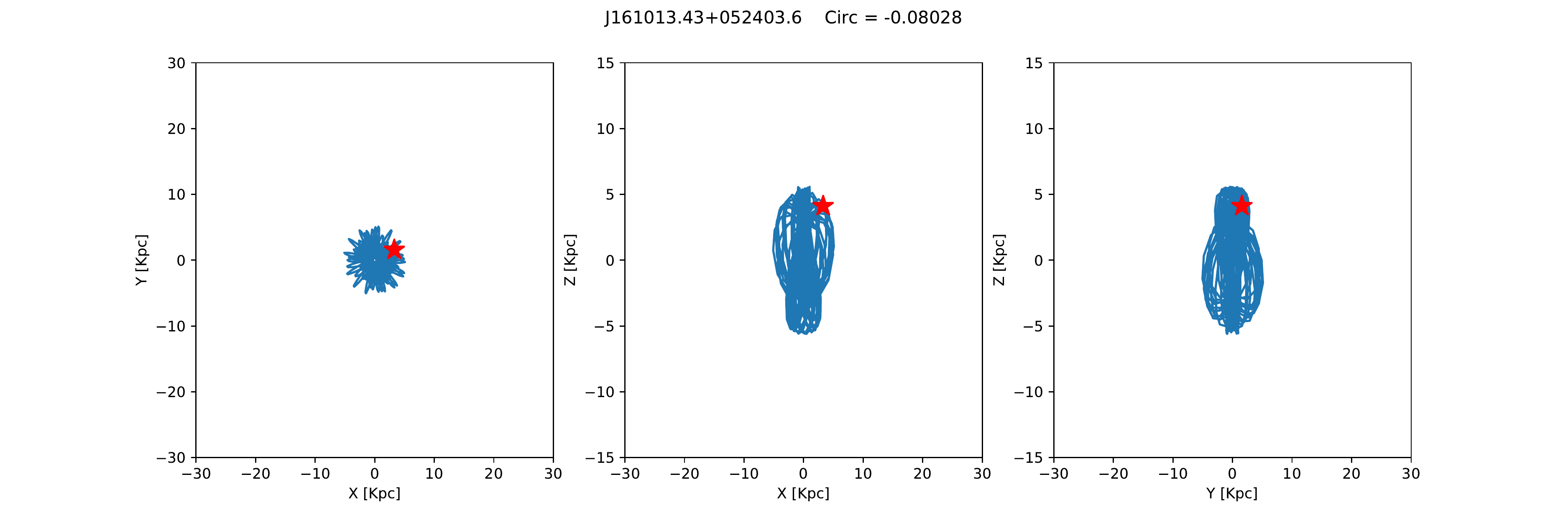}} \quad
        
    \caption{Continued.}
\end{figure*}

\begin{figure*}\ContinuedFloat
\captionsetup[subfigure]{labelformat=empty}
        \subfloat[][]
        {\includegraphics[width=\textwidth]{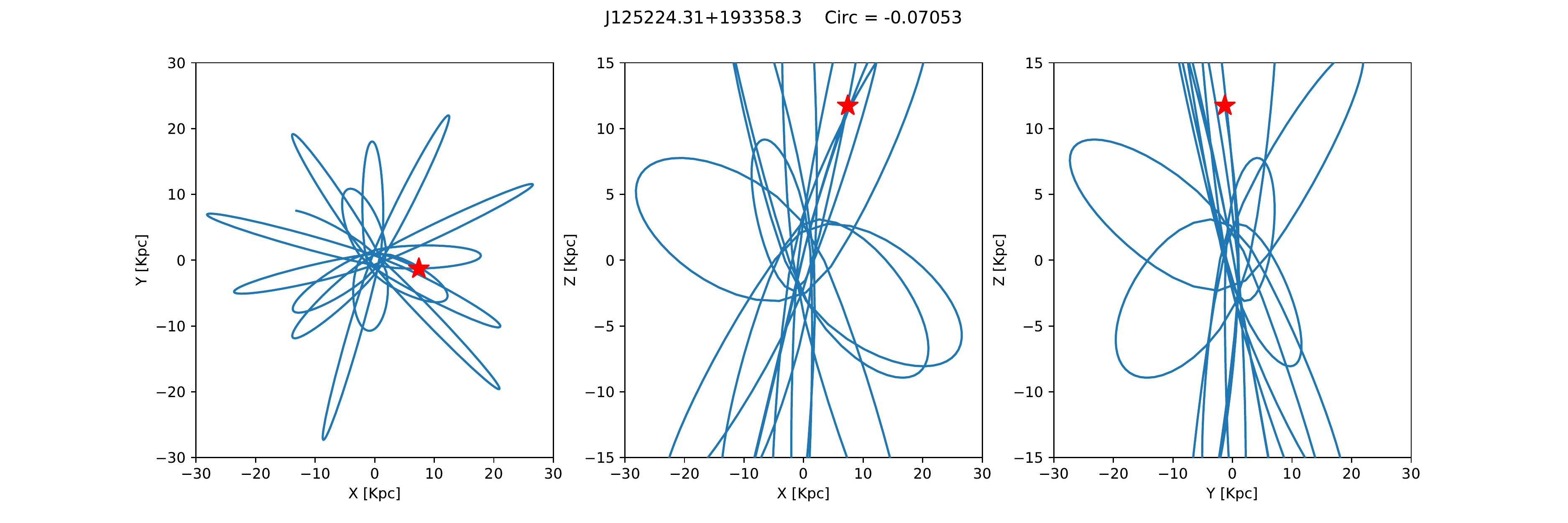}} \quad
            \subfloat[][]
        {\includegraphics[width=\textwidth]{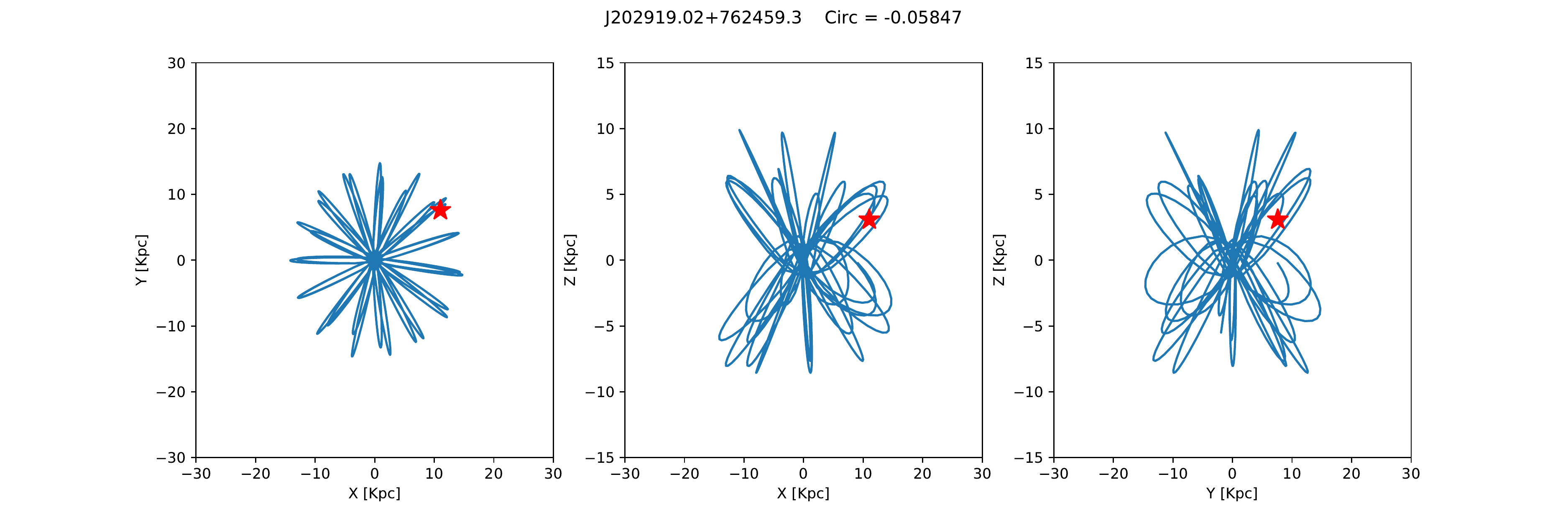}} \quad
            \subfloat[][]
        {\includegraphics[width=\textwidth]{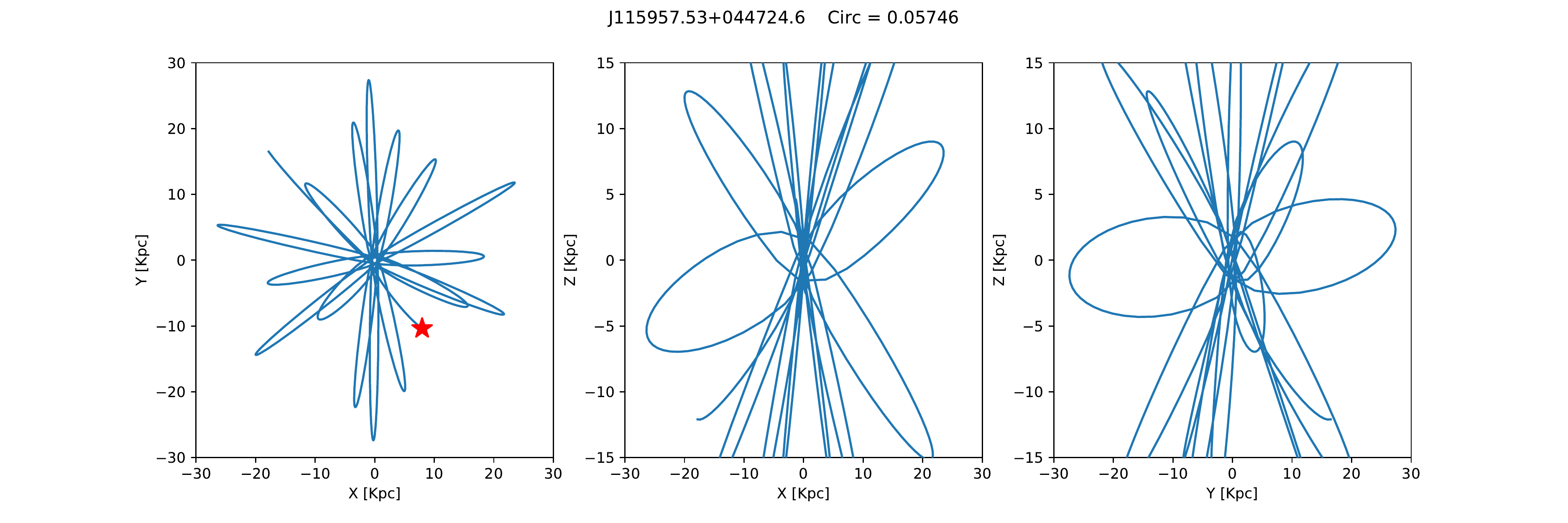}} \quad
        
    \caption{Continued.}
\end{figure*}

\begin{figure*}\ContinuedFloat
\captionsetup[subfigure]{labelformat=empty}
        \subfloat[][]
        {\includegraphics[width=\textwidth]{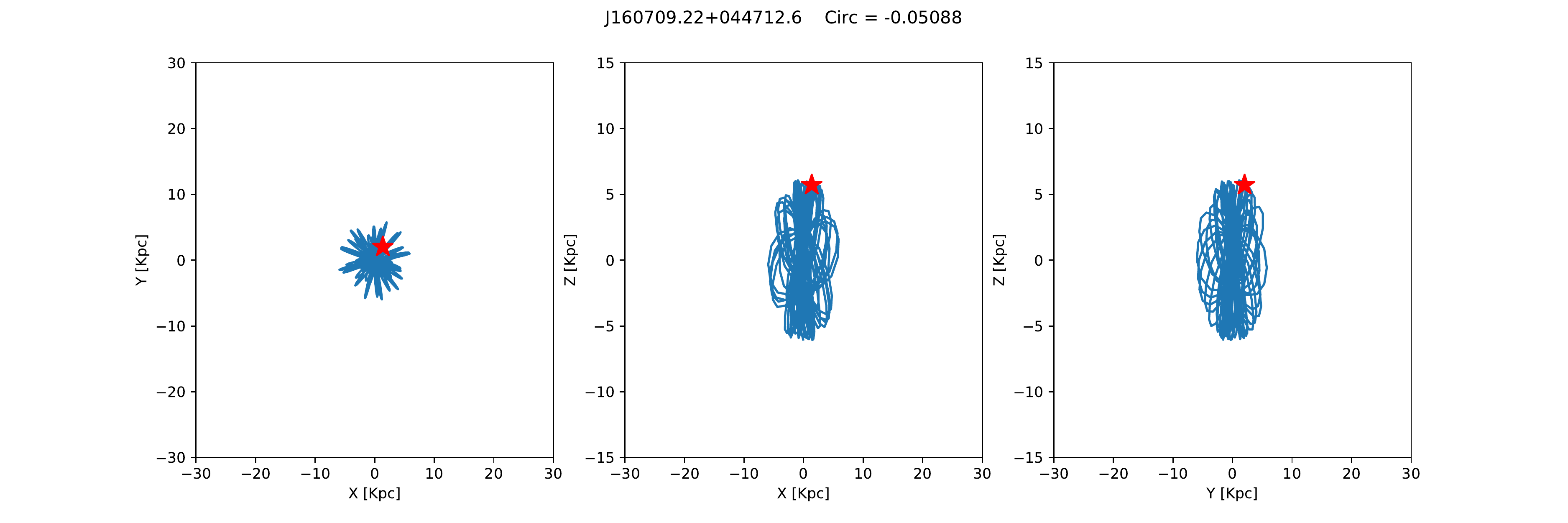}} \quad
            \subfloat[][]
        {\includegraphics[width=\textwidth]{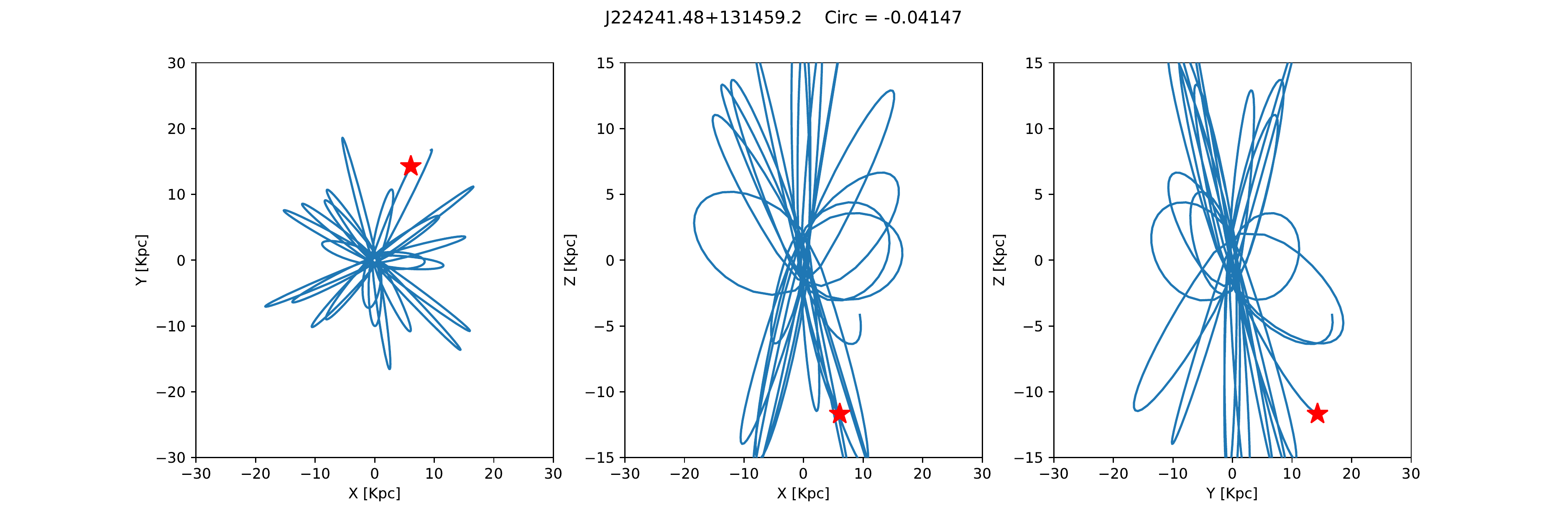}} \quad
            \subfloat[][]
        {\includegraphics[width=\textwidth]{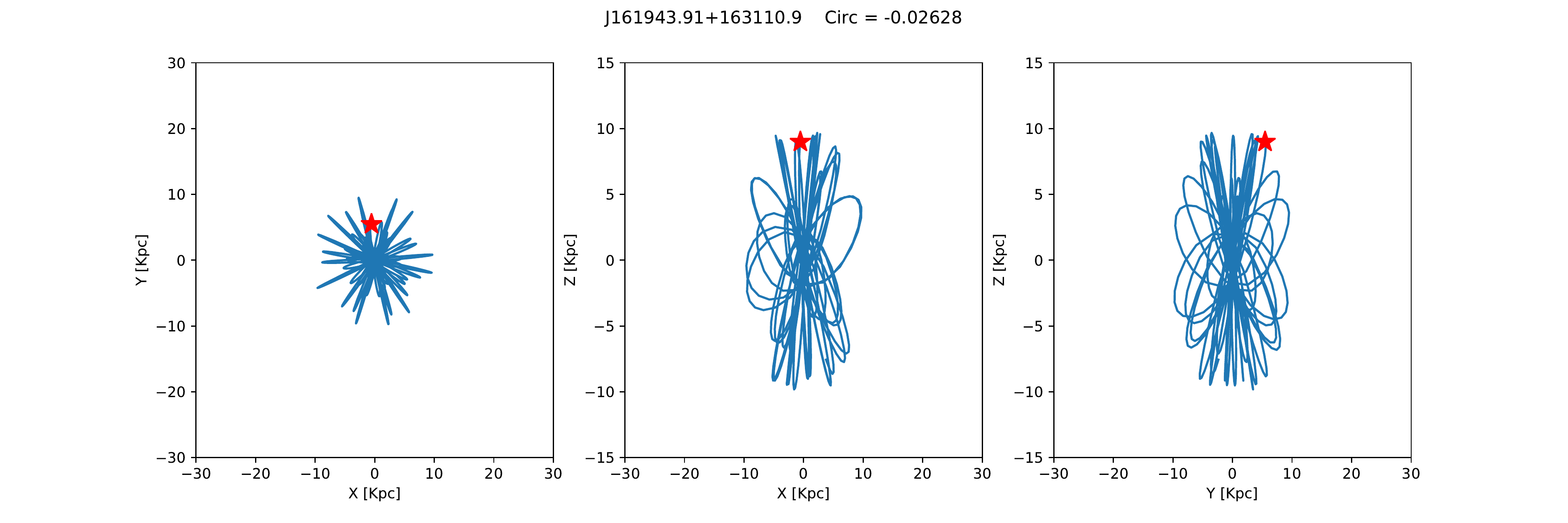}} \quad
        
    \caption{Continued.}
\end{figure*}
\end{document}